\documentclass[fleqn,usenatbib]{mnras}

\usepackage{newtxtext,newtxmath}

\usepackage[T1]{fontenc}
\usepackage{ae,aecompl}

\usepackage{graphicx}	
\usepackage{subcaption}
\usepackage{CJKutf8}
\usepackage{amsmath}	
\usepackage{bm}
\usepackage{multirow}
\usepackage{threeparttable}
\usepackage{float}
\usepackage{epsf}
\usepackage{booktabs}
\usepackage{mathrsfs} 

\newcommand{\msunh}{\>h^{-1}\rm M_\odot}

\newcommand{\Msun}{\>{\rm M_{\odot}}}

\newcommand{\mpch}{\>h^{-1}{\rm {Mpc}}}

\newcommand{\rmH}{{\rm H}}
\newcommand{\rmf}{{\rm f}}
\newcommand{\rmg}{{\rm g}}

\newcommand{\rmA}{{\rm A}}

\title[HIR4: HI-galaxy cross correlation]{HIR4: cosmological signatures imprinted on the cross- correlation between a 21-cm map and galaxy clustering}

\author[Shi et al]{\parbox{\textwidth}{
Feng Shi$^1$\thanks{E-mailich:fengshi@kasi.re.kr}, 
Yong-Seon Song$^1$,
Jacobo Asorey$^{1,2}$,
David Parkinson$^1$,
Kyungjin Ahn$^3$,
Jian Yao$^5$, 
Le Zhang$^{4,5}$ 
and Shifan Zuo$^{6,7}$
}\\\\
$^1$ Korea Astronomy and Space Science Institute, Yuseong-gu, Daedeok-daero 776, Daejeon 34055, Korea\\
$^2$ Centro de Investigaciones Energ\'eticas, Medioambientales y Tecnol\'ogicas (CIEMAT), Av. Complutense, 40, 28040 Madrid, Spain \\
$^3$ Department of Earth Sciences, Chosun University, Gwangju 61452, Korea\\
$^4$ School of Physics and Astronomy, Sun Yat-Sen University, 2 Daxue Road, Tangjia, Zhuhai, 519082, P.R. China\\
$^5$ Department of Astronomy, Shanghai Jiao Tong University, Shanghai, 200240, P.R. China\\
$^6$ Key Laboratory of Computational Astrophysics, National Astronomical Observatories, Chinese Academy of Sciences,\\
Beijing 100101, P. R. China\\
$^7$ Department of Astronomy and Tsinghua Center for Astrophysics, Tsinghua University, Beijing 100084, P. R. China
}

\date{Accepted XXX. Received YYY; in original form ZZZ}

\pubyear{2019}

\begin{document}
\label{firstpage}
\pagerange{\pageref{firstpage}-\pageref{lastpage}}
\maketitle

\begin{abstract}
We explore the cosmological multitracer synergies between an emission line galaxy distribution from the Dark Energy Spectroscopic Instrument and a Tianlai Project 21-cm intensity map. We use simulated maps generated from a particle simulation in the light-cone volume (Horizon Run 4), sky-trimmed and including the effects of foreground contamination, its removal and instrument noise. We first validate how the foreground residual affects the recovered 21-cm signal by putting different levels of foreground contamination into the 21-cm maps. We find that the contamination cannot be ignored in the angular autocorrelation power spectra of HI even when it is small, but has no influence on the accuracy of the angular cross-power spectra between HI and galaxies. In the foreground-cleaned map case, as information is lost in the cleaning procedure, there is also a bias in the cross-correlation power spectrum. However, we found that the bias from the cross-correlation power spectrum is scale-independent, which is easily parameterized as part of the model, while the offset in the HI autocorrelation power spectrum is non-linear. In particular, we tested that the cross-correlation power also benefits from the cancellation of the bias in the power spectrum measurement that is induced by the instrument noise, which changes the shape of the autocorrelation power spectra but leaves the cross-correlation power unaffected. We then modelled the angular cross-correlation power spectra to fit the baryon acoustic oscillation feature in the broad-band shape of the angular cross-correlation power spectrum, including contamination from the residual foreground and the effect of instrument noise. We forecast a constraint on the angular diameter distance $D_\mathrm{A}$ for the Tianlai Pathfinder redshift $0.775<z<1.03$, giving a distance measurement with a precision of 2.7\%  at that redshift. 
\end{abstract}




\section{Introduction}
Understanding the nature of the accelerated expansion of the Universe is one of the most important currently outstanding problems in cosmology. In the last few decades, cosmological observations from galaxy surveys, the cosmic microwave background (CMB), Baryon Acoustic Oscillations (BAOs), weak gravitational lensing shear (WL), gravitational wave standard candles, and other observations, have all made significant contributions to our understanding of the Universe. However, there is still much uncertainty in our knowledge about dark energy and the physics of the expansion of the Universe, such as the $H_0$ tension between the CMB and low- redshift measurement \citep{2016JCAP...10..019B}, and the varying dark energy \citep{2017NatAs...1..627Z}, which requires further tests. In fact, the intermediate time line of our cosmic distances and expansion rates from observations has not yet been systematically surveyed.

The 21-cm intensity mapping is a technique for surveying the large-scale structure of the Universe, by using the 21-cm spectral line that arises from the hyperfine `spin flip' transition of neutral hydrogen (HI). It measures the integrated emission lines that originate from many unresolved galaxies that trace the HI gas, which follows fluctuations in the underlying cosmic density field. As the frequency of the emission line is redshifted by the expansion of the Universe, it is possible to detect the underlying clustering signal as a function of redshift. This is in principle similar to the traditional galaxy redshift survey, but with an important distinction that 21-cm intensity mapping is sensitive to all sources of emission, rather than just cataloguing the brightest galaxies. As high angular resolution is not required, 21-cm intensity mapping can cover large sky areas in a limited observing time. It can explore larger volumes and measure BAOs continuously from low to high redshift. For example, CHIME \citep{2014SPIE.9145E..4VN}, the Tianlai project \citep{2011SSPMA..41.1358C, 2015ApJ...798...40X, 2018SPIE10708E..36D}, the SKA \citep{2015aska.confE..19S} and the Hydrogen Intensity and Real-time Analysis eXperiment (HIRAX) \citep{2016SPIE.9906E..5XN} are currently being built and taking data with the goal of measuring the BAO scale to $z=2.5$ with unprecedented precision. This is ideal for testing the time variation of the dark energy equation of state \citep{2018arXiv180411137D}, and the  problem of the Hubble parameter inconsistency between CMB and local measurements, with a single tracer across a quite wide redshift range \citep{2019BAAS...51c.101K}.


However, a key problem of the endeavor of such a successful measurement is that 21-cm signal is contaminated by several sources of foreground radiation, which are many orders of magnitude brighter than the cosmological signal. Great efforts have been make to study the foreground removal, using, for example, principal component analysis \citep[PCA,][]{2008MNRAS.388..247D} and fast independent component analysis \citep[FastICA,][]{2012MNRAS.423.2518C,FastICA} method. However, because the foreground noise is tangled with the 21-cm signal, there would be always residual foreground on the map, together with the signal information loss from the cleaning procedure. 

One possible way to mitigate the contamination by the residuals is through measurement of the cross-correlation. So far, the detection of the cross-correlation between the large-scale structure and 21-cm intensity maps has been reported in \citet{2010Natur.466..463C}, based on the data from the Green Bank Telescope and DEEP2 galaxy survey at $z \sim 0.8$, and in \citet{2018MNRAS.476.3382A}, based on the data from the Parkes radio telescope and the 2dF galaxy survey at $0.057 < z < 0.098$. Based on the simulation data, great efforts have been made not only to understand the HI clustering from autocorrelation or cross-correlation,  but also to extract cosmological information \citep[e.g.][]{2015ApJ...798...40X,2019MNRAS.488.5452C,2019MNRAS.485.5519W,2019arXiv190911104P,2020MNRAS.493.5854H,2020arXiv200205626C}.

In order to be more realistic, our HIR4 (HI with Horizon Run 4) project aims to investigate the prospects for probing the HI clustering from  future 21-cm intensity maps in a manner closer to observations. In our first paper \citep{2020arXiv200100833A}, we simulated the future survey of 21-cm using the Horizon Run 4 (HR4) \citep{2015JKAS...48..213K} cosmological N-body simulation in the light cone volume. We generated HI intensity maps from the halo catalogue, and combined with foreground radio emission maps from the Global Sky Model, to create accurate simulations over the entire sky. We simulated the HI sky to match the sensitivity of the Tianlai Pathfinder. 

In this paper, we focus on testing how the residual foreground affects the clustering signal imprinted in the 21-cm intensity map, and validate the detectability of HI clustering from the 21-cm autocorrelation and 21-cm $\times$ galaxy cross-correlation angular power spectrum when using the foreground-cleaned 21-cm maps. We also forecast the angular diameter distance constraint from the cross correlation between the Dark Energy Spectroscopic Instrument (DESI) \citep{2016arXiv161100036D} galaxy survey and the Tianlai 21-cm intensity mapping. The Tianlai Project is designed to demonstrate the feasibility of using wide-field-view radio interferometers to map the density of neutral hydrogen, which is divided into three stages: Pathfinder, Pathfinder+ and Full Array. At present, construction of the Pathfinder has been completed and it is now undergoing the calibration process before the survey starts in the near future. The DESI is a new instrument for conducting a spectroscopic survey of distant galaxies, which achieved its first light test in October 2019, and survey validation with the completed instrument began in early 2020. Both surveys have a big overlap in sky coverage and redshift, which provides a great opportunity to measure the cross-correlation in the coming years.

This paper is organized as follows: In Section~\ref{sec:simu} we introduce the mock data to simulate the galaxy distribution and 21-cm intensity mapping including foreground contamination and instrument noise. In Section~\ref{sec:mea}, we test the detectability of HI clustering from the foreground-removed 21-cm intensity maps. Modelling the angular cross-correlation power spectra and fitting the broad-band BAO feature is presented in Section~\ref{sec:baofit}. Finally, we summarize our conclusions in Section~\ref{sec:conclu}.

%
\begin{figure*}
  \centering
  \includegraphics[width=2.0\columnwidth]{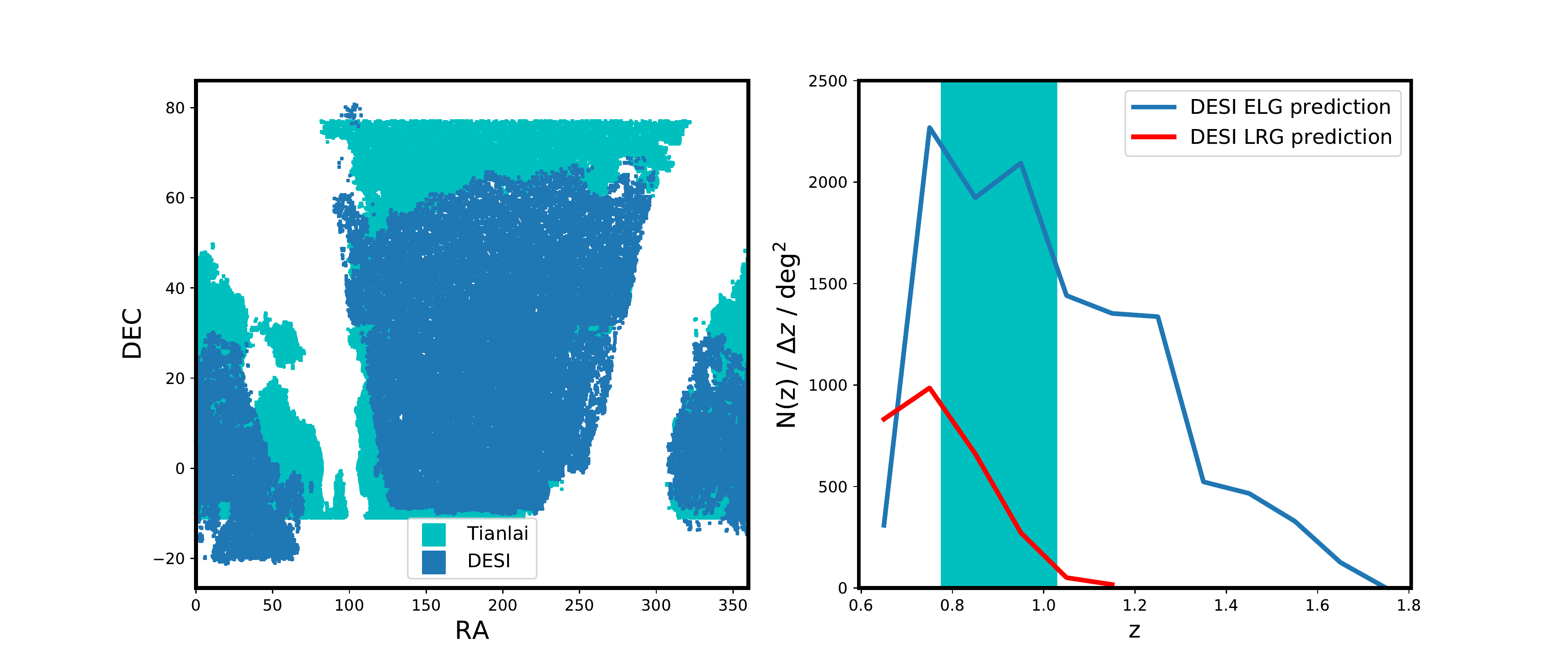}
  \caption{(Left panel) The target angular distribution of the ELG (blue) and Tianlai (cyan) after applying the corresponding mask scheme. (Right panel) The redshift distribution of the DESI expected number density of ELG (blue line) and LRG (red line), which is obtained from \citet{2016arXiv161100036D}. The cyan band corresponds to the redshift range of the Tianlai Pathfinder.}
  \label{fig:map_distri}
\end{figure*}
\section{Mock Catalogue}\label{sec:simu}
The deepest and widest large-scale structure experiments will commence in the near future, designed to observe various targets, such as luminous red galaxies, emission line galaxies, the shear or the magnification of galaxy morphology caused by gravitational lensing, and 21-cm emission of neutral atomic hydrogen HI. Both spectroscopic surveys and 21-cm emission observations on a large scale are planned to scan nearly the same region of sky over the northern hemisphere, and it will be possible to open a new window by cross-correlating them to reveal the underlying matter fluctuations in a different way. While each experiment suffers from  systematic uncertainties, such as the fibre assignment contamination for the spectroscopic survey and  foreground contamination for the 21-cm intensity mapping survey, the cross-correlated measurement is immune to those errors, and provides an independent representation of the invisible large-scale structure of cold dark matter. The detectability of underlying clustering by exploiting cross-correlating methodology is verified by analyzing the mock catalogue in which both targets are generated from the same realization of N-body simulation. 


\subsection{The planned surveys}

The DESI has been constructed to probe the signature of dark energy on the expansion of the Universe, by obtaining spectroscopy measurements of numerous galaxies to construct a huge three dimensional map, stretching from the local universe up to  a distance of 10 billion light years. Emission-Line Galaxies (hereafter ELG) are one of the major targets at $z>0.6$ for DESI. The strong nebular emission lines originate in the HII ionized areas where the short-lived luminous and massive stars are found. ELGs are generally late-type spiral and irregular galaxies, but a few galaxies with a large number of new born stars are accepted as belonging to the ELG type as well. The active star formation rate of ELGs means that the massive stars dominate the integrated rest-frame colors, and they appear bluer than Luminous Red Galaxies (hereafter LRGs).

ELGs are selected using optical color-selection techniques, slicing in optical color-color space to effectively isolate the population of $0.6 < z < 1.7$ galaxies, a method that has been confirmed by multiple experiments. A high redshift success rate is expected  for ELGs with  integrated OII emission-line strengths of $8\times 10^{-17} {\rm erg/s/cm^2}$, corresponding to a limiting star-formation rate of 1.5, 5 and 15 M$_{\odot}$/yr at $z\sim 0.6$, 1 and 1.6, respectively, which can be well identified at $z>0.6$ \citep{2016arXiv161100036D}. The separation of galaxies with redshift lower and higher than $z\sim 0.6$ is possible by the spectrum blue-ward of the Balmer break moving toward the z-band filter, which causes $g - r$ and $r - z$ colors to be relatively blue at higher redshifts. After applying those specifications, we estimate the ELG target distribution in redshift presented in the blue region in the left panel of Figure \ref{fig:map_distri}.

The Tianlai Project plans to demonstrate the feasibility of using wide field of view radio interferometers to map the density of neutral hydrogen in the northern sky after the epoch of reionization, which provides an inexpensive means for surveying the large-scale structure of the Universe. The program is constrained by the site location which should be free from all types of ground radio noises. It sits on Hongliuxia, a radio-quiet site in northwest China of $44^\circ 9'9.66''$ N $91^\circ 48'24.72''$ E. The construction schedule is divided into three stages - Pathfinder, Pathfinder+ and Full Array. Construction of the Pathfinder was completed in 2016 and it is now taking data on a regular basis. The Pathfinder instrument consists of both a dish array and a cylinder array. The large-scale survey will be carried by the cylindrical-type instrument which scans the North Celestial Cap as the first goal. The cylinder includes three adjacent cylindrical reflectors oriented in the North-South direction. Each of the cylinders is 15m wide and 40m long. Currently, the central 12.8m parts of the cylinders are equipped with receiver feeds, each having 31, 32, and 33 from East to West respectively. The target survey area is presented in the green region in the left panel of Figure \ref{fig:map_distri}, which is well overlapped with the DESI ELG target sky patch. When both are cross-correlated, most DESI targets observed at the northern hemisphere contribute to the cross-correlation computation.

The Tianlai Pathfinder will be operated at the frequency range of 700-800MHz, which corresponds to the redshift span of $0.78<z<1.03$, computed using HI emission frequency at the rest frame. According to the DESI predicted redshift distribution of number density \citep{2016arXiv161100036D}, shown in the right panel of Figure \ref{fig:map_distri}, DESI ELG targets have larger and more complete number density in this redshift range while DESI LRG targets are much less common at that high redshift.  It is therefore ideal to cross-correlate between  HI emission from Tianlai and the DESI ELG observations.
\begin{figure*}
    \centering
    \includegraphics[width=2.27\columnwidth]{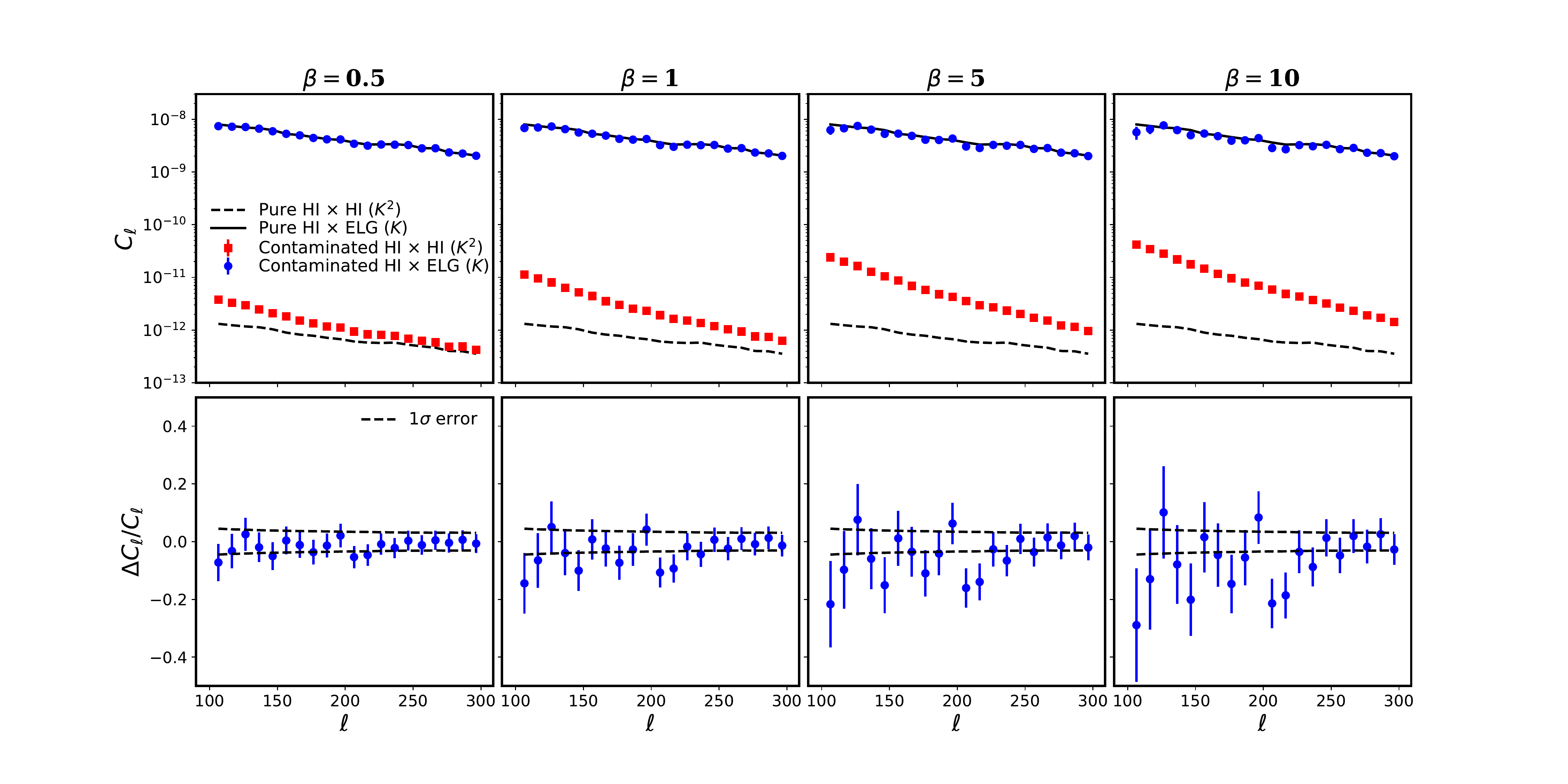}
    \caption{Toy model for testing the angular power spectrum using the 21-cm maps contaminated by different-level foregrounds. The contamination level is parameterized by the ratios of the temperature variance between the foreground and the 21-cm signal (see Equation~(\ref{eq:beta})). In the upper panels, the red squares and blue circles correspond to HI $\times$ HI and HI $\times$ ELG of realistic temperature map with mixture of HI and foreground noise. For comparison, the black dashed and solid lines show the corresponding pure power spectrum with no foreground noise on the map. Note that the pure power spectra are same in the different panels. In the lower panels, the blue circles show the normalized difference of $C_\ell$ between the contaminated and the pure HI $\times$ ELG. The black dashed lines indicate the normalized $1\sigma$ error range of the pure cross-correlation power spectrum. Note that the units for autocorrelation and cross-correlation $C_\ell$ are different, as indicated in the legend.}\label{fig:Cl_toymodel}
\end{figure*}
%
\subsection{Creating ELG and HI mock catalogues}
\label{sec:mocking}

The HR4 \citep{2015JKAS...48..213K} simulation is used to generate both HI intensity mapping and galaxy distribution mocks. This particle mock is a high-resolution N-body simulation that evolves the distribution of $6300^3$ dark matter particles in a periodic cubic box of $3150 \mpch$ on a side. It adopted a standard $\Lambda{\rm CDM}$ cosmology in concordance with WMAP 5-year results, and each particle has a mass of $9\times 10^9\Msun$. Specifically, the matter, baryonic matter, and dark energy densities are $\Omega_{\rm m}=0.26$, $\Omega_{\rm b}=0.044$, and $\Omega_\Lambda=0.74$, respectively. The current Hubble expansion is $H_0=72{\rm km~s^{-1} Mpc^{-1}}$ and the clustering amplitude of matter on scales of $8\mpch$ is $\sigma_8 = 1/1.26$. In addition to the box volume, the HR4 team also built a past light-cone space data of haloes that cover all the sky up to $z \simeq 1.5$. Below we present the way to make ELG galaxy and HI intensity mocks.

The statistics for the distribution of dark matter particles is assumed to be linked to those of galaxies. The statistical methods to match the observed galaxy properties to those of dark matter halos have been developed, and are quantitatively expressed by the halo mass function and the halo bias tracing. While galaxies are gravitationally bound to dark matter halos and their evolution is tightly correlated with host halo, there are plenty of missing elements which are not presented in the halo model alone. Some, but not all, galaxies exhibit star formation quenching to be passive, which causes a bimodal blue or red distribution in the galaxy population, with few found in between. Those galaxies increase with host halo mass and become more frequent at the present time. In this paper, we adapt the constraints on the relation between host dark matter halos and galaxies which are measured by observationa of the peak location of the stellar to halo mass ratio using the combination of CFHTLenS and VIPERS \citep{2015MNRAS.449.1352C}. Among central and satellite galaxies populated for a given host halo, we identify the satellite galaxies with their host halo masses above the threshold bound $10^{12} \msunh$ as DESI ELGs. The redshift profile is presented in the right panel of Figure \ref{fig:map_distri} (blue), which is made by trimming it to fit to the DESI forecast of the ELG target distribution.

The 21-cm maps have been created see our previous paper \citep{2020arXiv200100833A} based on the HR4 simulation. Here we briefly summarize the main ingredients of the generation method and refer the reader to the previous paper for more details. The HI distribution is simulated by assigning HI to the halo according to the halo model, which estimates the mass of neutral hydrogen from the host halo mass \citep{2015MNRAS.454..218B,2016MNRAS.458..781P, 2017MNRAS.464.4008P}. The 21-cm brightness temperature, $T_\rmH$, can then be modelled from the HI density field \citep{2013MNRAS.434.1239B,2015ApJ...803...21B}. In practice, we stack the hydrogen mass hosted by the corresponding halo mass in each cube defined by an angular pixel and a redshift bin. The corresponding hydrogen mass, $M_\rmH$, is used to generate the temperature maps. In addition, as the limitation of the HR4 halo mass resolution, the HI temperature would be below the expected one in nature.  We corrected for this by rescaling the temperatures to meet the observed $z-\bar{T}_{\rm 21-cm}$ relation. A suite of foreground maps for each frequency bin of our mock catalogues have to be added, for which we used the Global Sky Model (hereafter GSM) \cite{2008MNRAS.388..247D,2017MNRAS.464.3486Z}. The GSM model maps include information from five different foregrounds: synchrotron, free-free, CMB, warm dust, and cold dust. 

%
\begin{figure*}
    \centering
    \includegraphics[width=0.9\columnwidth]{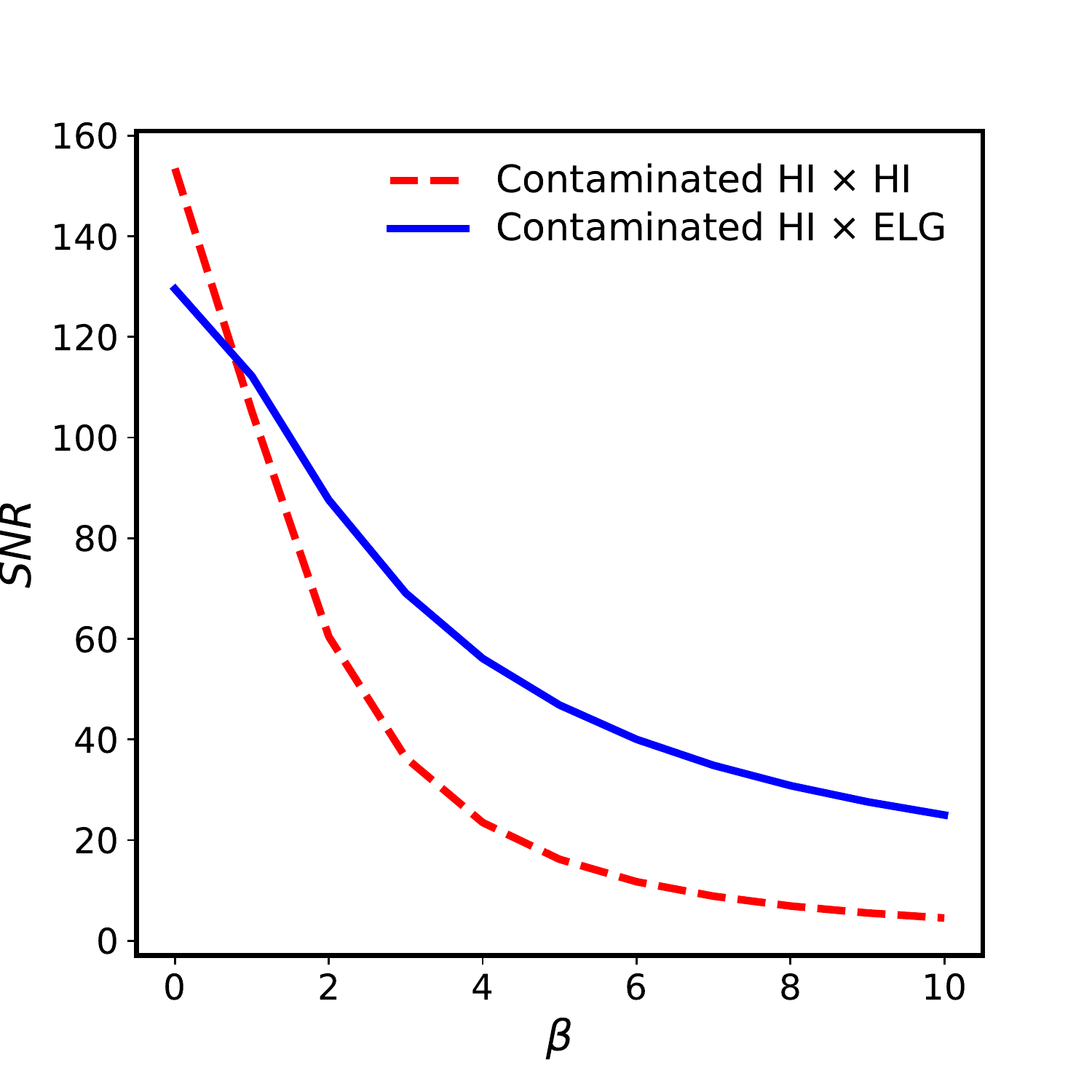}
     \includegraphics[width=0.9\columnwidth]{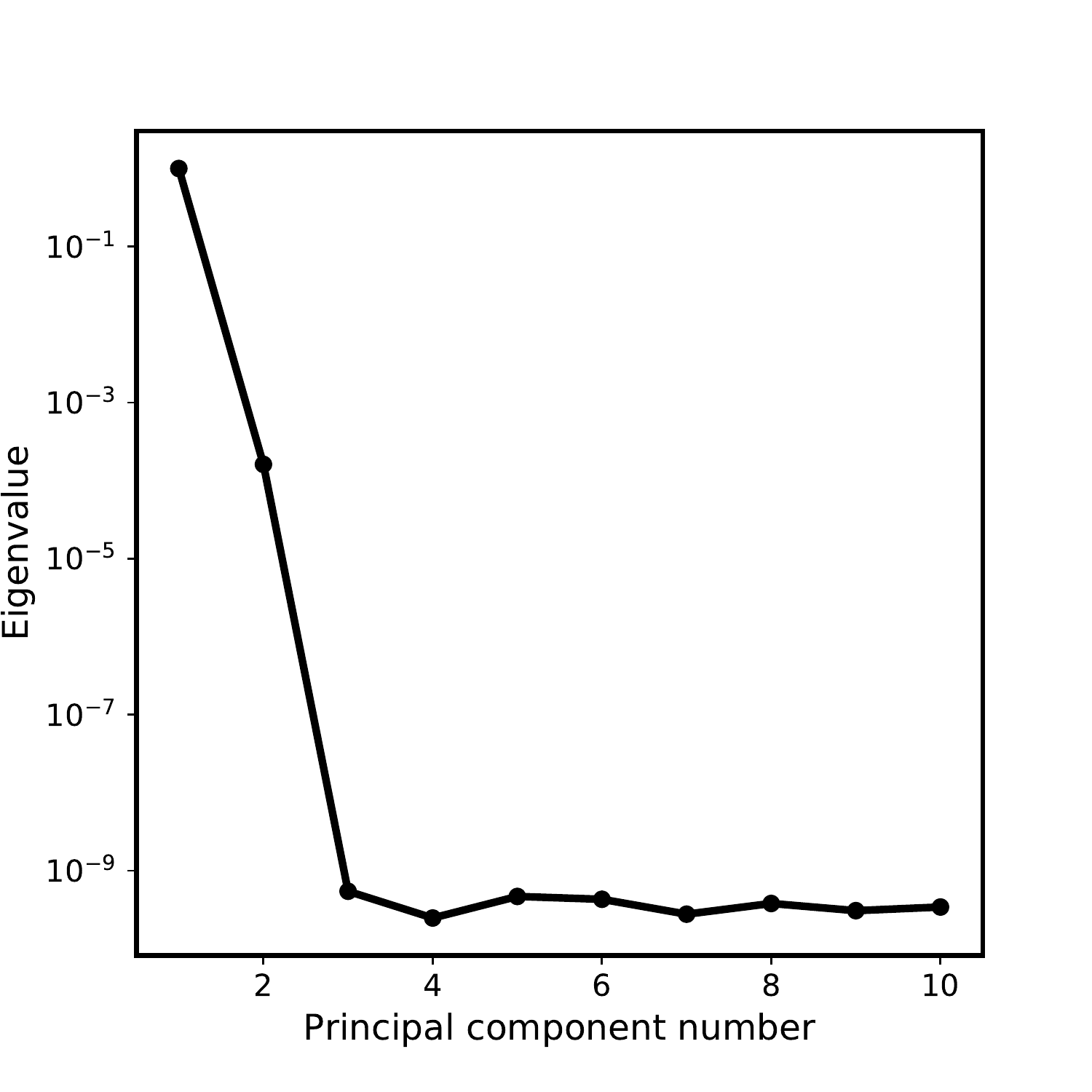}
    \caption{({\it Left panel}) Toy model for the SNR as a function of $\beta$. The red dashed line is shown for the results of the contaminated HI $\times$ HI. The blue solid line is for the contaminated HI $\times$ ELG. ({\it Right panel}) The eigenvalues of the 10 principal components in the PCA foreground cleaning.}
    \label{fig:SNR_toymodel}
\end{figure*}

Finally, the instrument noise is estimated by assuming uncorrelated thermal noise across all baselines and frequencies. The noise level is represented in units of brightness temperature given by \citep{2001isra.book}:
\begin{equation}\label{eq:noise_cylinder}
\sigma^N_{ij}= \left({\bf N}_m^{ij}\right)^{1/2}= \frac{T_{\rm sys}}{\sqrt{\Delta\nu \Delta t_{ij}}}\left(\frac{\lambda^2}{A_e} \right)\,,
\end{equation}
where $\Delta t_{ij}$ is the total integration time of baseline $ij$, $T_{\rm sys}$ is the system temperature, $A_e$ is the effective area of the antenna, $\lambda$ is the observing frequency, and $\Delta\nu$ is the width of the frequency channel. The system temperature is the sum of the sky brightness and the analog receiver noise temperature, $T_{\rm sys} = T_{\rm sky} + T_{\rm rec}$. At the frequency of interest, the Tianlai array would be expected to achieve a total system temperature of $50-100$ K, and thus we assume $T_{\rm sys}$ = 50K in this study. We also assume two full years of observation for the Tianlai pathfinder survey. The effective antenna area $A_{e}$ is calculated by $A_e\Omega = \lambda^2$, where the beam solid angle $\Omega$ is well approximated by $\Omega\simeq 0.1$ for the current Tianlai cylinder array. 

Now the observed temperature is given as the one that combines the cosmological signal from the simulation with the foregrounds and the observational noise,
\begin{equation}
    T^{\mathrm{obs}}_{b} (\hat{n}) = T^{\mathrm{HI}}_{b}(\hat{n}) + T^{\mathrm{foreground}}_{b}(\hat{n}) + T^{\mathrm{noise}}_{b}(\hat{n}).
\end{equation}

The foreground signal computed using the GSM is much bigger than the cosmological signal $T_{21}$, and so the two cannot be simply decomposed. Instead, reconstruction methods are developed to split all different types of foreground emissions caused by diverse origins. The local signal and that of cosmological origin will have distinguishable frequency dependence, which can be reconstructed using, for example, methods such as fast independent component analysis (FastICA), principal component analysis (hereafter PCA), and log-polynomial fitting. In this paper, we exploit the PCA and FastICA methods to decompose the foreground noise from the cosmological signals.

Our descriptions of these two methods match those in \citet{2020arXiv200100833A}, but we summarize again here:

\noindent PCA: The pre-whitening method is applied to subtract the mean of simulated data. The covariance matrix of the data, after the pre-whitening procedure, is computed among different data bins in frequency space. This matrix in frequency space is decomposed into eigenvectors with distinct eigenvalues. It is assumed that the foregrounds dominate the eigenmodes with the highest eigenvalues, as those are the highest amplitude components of the power in the maps. It is observed that most foreground power exhibits a smooth curve in frequency space that is normally described by a combination of a few leading eigenmodes. In the PCA approach, those leading principal components with the largest eigenvalues in frequency space are projected from every spatial pixel to obtain foreground cleaned maps. There is a caveat that some correlations are cosmologically introduced as well, despite the smallness of the cosmological signal, which causes this cleaning process to affect the 21-cm signal slightly.

\noindent fastICA: The fastICA process assumes that the maps ($\mathbf{x}$) can decomposed into a set of independent signals $s_i$, through $\mathbf{x}=\mathbf{As}$ (with mixing matrix $\mathbf{A}$). This is done by maximizing the non-Gaussianity of the Independent Components, which are represented by the IC maps, and should correspond to the foregrounds. In contrast, the intensity of the cosmological 21-cm emission depends on the mass of neutral hydrogen present in the `voxel', which is a stochastic quantity with a Gaussian distribution, and so resembles the noise in a fastICA reconstruction process. Using the implementation of fastICA as part of the {\tt scikit-learn} python machine learning package \citep{scikit-learn}, we maximise the negentropy, defined by $J(y)=H(y_{\mathrm{gauss}})-H(y)$, assuming the negentropy is approximated by a $\log\cosh(y)$ function. As a measure of distance from gaussianity for the negentropy functions, maximizing it with respect to the components should remove the foreground signal, leaving behind the Gaussian cosmological signal.

\section{Large-scale HI distribution}
\label{sec:mea}

The mock HI signal embedded on large-scale structure is probed more efficiently using cross-correlation with a galaxy sample, over the same redshift range. We describe measuring the clustering using the angular power spectrum, and we investigate how well the HI clustering signal can be recovered from the foreground-cleaned map, and focus on making a comparison between autocorrelation and cross-correlation power spectrum. 

\subsection{Angular power spectrum}

For the galaxy, distribution, we use number density field, $n_{\rmg}$, where resolved galaxies can be counted in pixels within a redshift bin, and then we calculate the over-density, $\delta_\rmg$, as,
\begin{equation}
    \delta_\rmg(\alpha,\delta) = \frac{n_{\rmg}(\alpha,\delta)}{\bar{n}_{\rmg}}-1.
\end{equation}
Here $\alpha$ and $\delta$ are respectively the right ascension and declination in the equatorial coordinate system, and $\bar{n}_{\rmg}$ is the averaged galaxy number value over the map. For the 21-cm maps, we compute the temperature fluctuations as,
\begin{equation}
    \delta T_\rmH(\alpha,\delta) = T_\rmH(\alpha,\delta)-\bar{T}_\rmH.
\end{equation}
Next, we measure the angular power spectrum by decomposing the fluctuations into spherical harmonics in this way,
\begin{equation}
    \delta(\hat{n}) = 
     \sum_{\ell=0}^\infty
     \sum_{m=-\ell}^{m=\ell}{a_{\ell m}Y^m_\ell(\hat{n})}.
\end{equation}
The harmonics coefficients $a_{\ell m}$ describe the amplitude of the fluctuations in spherical harmonics space. Note that $\delta(\hat{n})$ corresponds to $\delta_\rmg(\alpha,\delta)$ or $\delta T_\rmH(\alpha,\delta)$. Then, the angular power spectrum is calculated by,
\begin{equation}\label{eq:Cl}
    C_\ell^{XY}=
    \frac{1}{2\ell+1}
    \sum_{m=-\ell}^{\ell}
    \left|a^{X}_{\ell m}a^{Y}_{\ell m}\right|^2,
\end{equation}
where $X$ and $Y$ denote tracers, such as `g' and `H' corresponding to the galaxy and HI, respectively. Meanwhile, the error can be estimated by,
\begin{equation}\label{eq:Cl_er}
    \Delta C_\ell^{XY} = 
    \sqrt{\frac{1}{(2\ell+1)\Delta \ell f_{\rm sky}}}
    \left[(C_\ell^{XY})^2 + C_\ell^{XX}C_\ell^{YY}\right]^{1/2},
\end{equation}
where $\Delta\ell$ is the bin width of $\ell$, and $f_{\rm sky}$ is the total observed sky fraction. Note that for the autocorrelation power spectrum, the error equation is reduced to
\begin{equation}\label{eq:Cl_er_auto}
    \Delta C_\ell^{XX} = 
    \sqrt{\frac{2}{(2\ell+1)\Delta \ell f_{\rm sky}}}C_\ell^{XX},
\end{equation}
which is verified by computing the dispersion of the corresponding azimuthal modes with the given $\ell$.
To measure the angular power spectrum we used the {\tt NaMaster}\footnote{Downloaded from \url{https://github.com/LSSTDESC/NaMaster}.} code \citep{NaMaster} with $\Delta\ell=25$, which uses the pseudo-$C_\ell$ (also known as MASTER) approach including the effect of the sky mask. In the measurements, we also find that our results are essentially insensitive to the bin width for reasonable choices.
\begin{figure*}
  \centering
  \includegraphics[width=2.27\columnwidth]{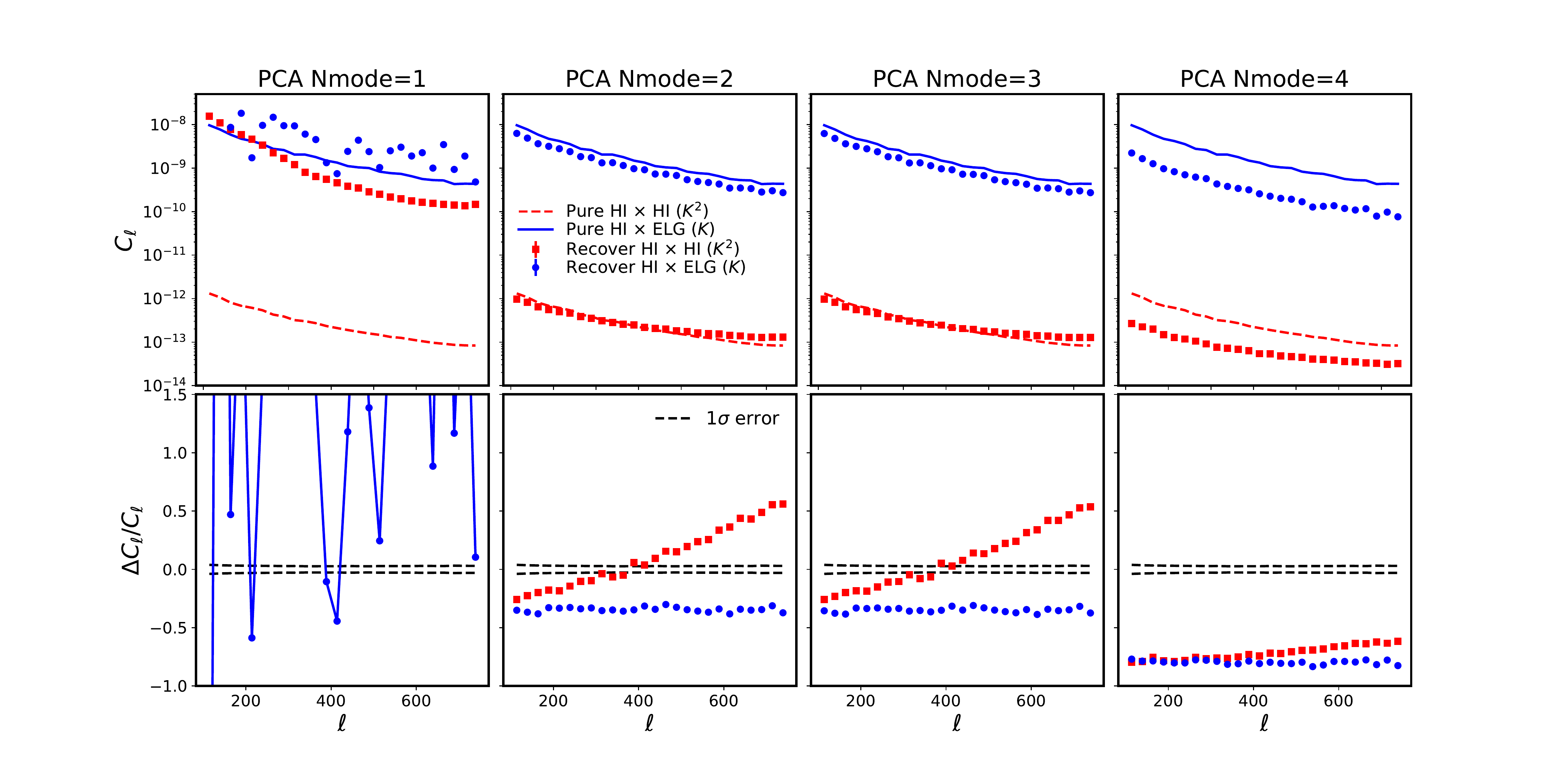}
  \caption{Comparison of the angular power spectra among the foreground-cleaning maps with a different number of PCA  modes removed. (Upper panels) Angular power spectrum for the $f=790-800$MHz frequency bin, with the reconstructed 21-cm map obtained by the PCA method. Different panels correspond to a different number of principal components removed, as the number indicated in the title of each panel. The red dashed and blue solid lines correspond to the true HI $\times$ HI and HI $\times$ ELG without contamination by foreground noise. The red and blue symbols are for the recovered HI $\times$ HI and HI $\times$ ELG, respectively. (Lower panels) Normalized difference between the recovered and the true power spectrum. The black dashed lines show the $1\sigma$ error ranges based on the pure HI $\times$ ELG. Note that the units for autocorrelation and cross-correlation $C_\ell$ are different, as indicated in the legend.}
  \label{fig:Cl_PCA_nmodes}
\end{figure*}
\begin{figure*}
  \centering
  \includegraphics[width=2.27\columnwidth]{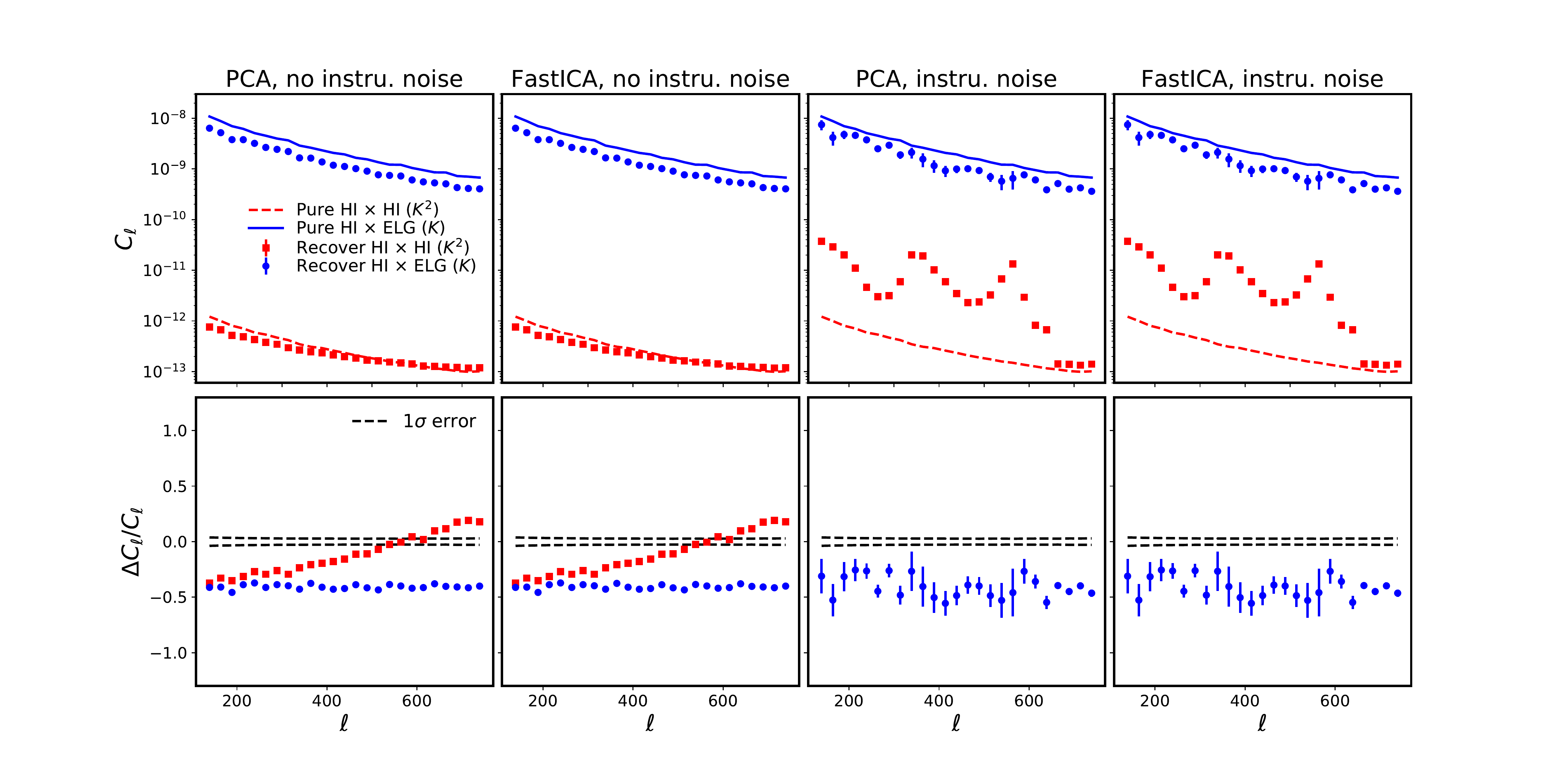}
  \caption{The measured angular power spectrum  as a function of multipole number $\ell$ from the Tianlai simulated and reconstructed 21-cm maps with two modes removed. The first two panels are measured from maps that do not include the Tianlai instrument noise, while the third and fourth panels do include it. (Upper panels) Angular power spectrum in the $f=700-710$MHz frequency bin. The reconstructed 21-cm maps are obtained by two different methods, PCA and FastICA, as indicated on the top. The red dashed and blue solid lines correspond to the true HI $\times$ HI and HI $\times$ ELG without any noise contaminated. The red and blue symbols are for the recovered HI $\times$ HI and HI $\times$ ELG, respectively. (Lower panels) Normalized difference between the recovered and the true power spectrum (where the true power spectrum is measured from the 21-cm without any noise or foregrounds). The black dashed lines show the $1\sigma$ error ranges based on the pure HI $\times$ ELG. Note that the units for autocorrelation and cross-correlation $C_\ell$ are different, as indicated in the legend.}
  \label{fig:Cl_PCA_FastICA}
\end{figure*}

\subsection{Threshold limit for residual foreground noise}
\label{sec:toymodel}

The foreground signal dominates over the HI temperature by five orders of magnitude, rendering the 21-cm clustering signal almost undetectable. Those foreground sources can be cleaned to a level that is close to the HI signal. However, there would be HI signal loss and residual foreground signal in the map, as those sources are tangled with the 21-cm signal in a nearly inseparable manner. Before applying the foreground removal method for the mock maps, we would like to validate how the foreground noise affects the 21-cm clustering signal, and to seek a guideline threshold level for the cleaning procedure. 

The residual foreground noises are assumed to be remaining mixed with the HI signal. The HI mock maps with full foreground noises is repainted with the sub-level residual noises. The sub-level noise maps are obtained by re-scaling the foreground map's temperature obtained from the GSM model (see details in Section \ref{sec:mocking}) to the desired contamination level, which is parameterized by the ratios of the temperature variance between foreground and 21-cm signal as,
\begin{equation}\label{eq:beta}
    \beta = \frac{\langle(T_\rmf-\overline{T}_\rmf)^2\rangle}
                 {\langle(T_\rmH-\overline{T}_\rmH)^2\rangle}.
\end{equation}
Note that $\beta=0$ corresponds to the pure HI map with no foreground noises. Figure~\ref{fig:Cl_toymodel} shows the comparison between the autocorrelation of HI $\times$ HI and the cross-correlation of HI $\times$ ELG. Here color symbols represent $C_\ell$ power spectrum of a realistic temperature map with a mixture of HI and foreground noise, and black curves represent the case with no foreground noise limit. Here we have four different-level contaminated maps, corresponding to $\beta=[0.5$, $1$, $5$, $10]$, as indicated in different columns. Note that the power spectra with no foreground noise are the same in the different panels. 

The HI $\times$ HI power spectra with foreground noise exhibits the biased result against the power spectrum at no foreground noise limit. Although the foreground noise is mixed with much smaller sub-level than HI signal of $\beta=0.5$, the contamination is cannot be ignored at large scale. The effect of the non-trivial autocorrelation of foreground noise is visible so we are not able to access to the underlying structure formation of particles. 

The cross-correlation power spectra of HI $\times$ ELG exhibit a better performance to probe HI signal than auto HI $\times$ HI spectra, with the remaining residual foreground contamination. While the foreground noise is strongly self-correlated in HI $\times$ HI, it does not correlate with the galaxy sample collected at much higher redshift in which the HI signal is radiated. Because foreground noise does not trace the underlying particle density fields which both HI signal and galaxies do, no contribution of the cross term between foreground noise and galaxy is presented in HI $\times$ ELG. To see precisely the performance, we plot the normalized difference of $C_\ell$ between the contaminated and pure HI $\times$ ELG in the lower panels of Figure~\ref{fig:Cl_toymodel}. The black lines indicate the normalized $1\sigma$ error range of the pure cross-correlation power spectrum, where error is computed by equation~(\ref{eq:Cl_er}). We can see that the contaminated is consistent to the pure within the $1\sigma$ level in the $\beta=0.5$ and $1$ cases. In the $\beta=5$ and $10$ cases, although the contaminated fluctuate outside the $1\sigma$ level, there is no obvious bias to the pure cross. This indicates that, in the case of HI $\times$ galaxy cross-correlation, the foreground contamination could have no influence on the accuracy but affects the precision only.

In order to evaluate the precision influenced by the foreground in a further step, we compute the signal-to-noise ratio (SNR) as a function of $\beta$ by,
\begin{equation}
\label{eq:SNR}
    {SNR}(\beta) = 
    \sqrt{ 
         \sum^{\ell_{\rm max}}_{\ell_{\rm min}}
         \left(\frac{C_\ell(\beta=0)}{\Delta C_\ell(\beta)}\right)^2
         },
\end{equation}
where $C_\ell (\beta=0)$ is the power spectra measured from the HI map with no foreground noise, $\Delta C_\ell (\beta)$ is the error of the power spectra measured from the HI map contaminated by $\beta$-level foreground noise, and the summation is computed from $\ell_{\rm min}=0$ to $\ell_{\rm max}=300$. In the left panel of Figure~\ref{fig:SNR_toymodel}, we plot the SNR as a function of $\beta$ for both of the HI $\times$ HI (dashed) and the HI $\times$ ELG (solid) power spectra. As expected, HI $\times$ ELG retains a larger SNR than the HI $\times$ HI after around $\beta=1$, as the cross correlation could remove most of the non-correlated noise and keep more information inside. Meanwhile, we find that the SNR is going down rapidly when increasing $\beta$ from $1$ to $5$. This indicates that it should be possible to significantly increase the sensitivity of detecting HI clustering when cleaning the foreground to the $\beta=1$ level.

\subsection{Foreground cleaning}

The threshold limit of the mixture level of foreground noise was presented in the previous subsection. We have found that the HI signal is accurately measured even with the residual foreground noise remaining around $\beta\sim 1$, but only when using the HI $\times$ ELG cross-correlation. It is plausible that one might be developed, but presently there is no perfectly selective foreground cleaning that will leave no damage to the HI signal. The information of HI signal and foreground noise is entangled in a manner not to be perfectly separable. It is expected that foreground cleaning to reduce $\beta$ will cause some significant damage to the HI signal. In this subsection, practical cleaning methods such as PCA and FastICA are applied to the simulated map, to investigate the level of information loss and the impact on the detectability of HI signal. In addition, the instrument noise is added to make the prediction more realistic.


We divided the 21-cm map into 10 frequency bins within the Tianlai frequency range, and applied the PCA method to remove foregrounds that are essentially smooth in the frequency direction and have very strong frequency coherence. The right panel of Figure~\ref{fig:SNR_toymodel} presents the 10 eigenvalues of frequency-frequency covariance matrix of the mock data. As shown, the first two modes capture more than 99 per cent of the information, dominating the whole sky. Such modes can be regarded as the foreground-dominated modes, so that their removal should mostly leave the foreground noise out and keep the 21-cm-dominated signal in maps with high fidelity. 

We also checked to make sure the foreground is removed at the appropriate level. We created maps of the remaining 21-cm distribution with different numbers of modes removed from the original map, and compare the recovered power spectrum to the cases with no foreground. Results are shown in Figure~\ref{fig:Cl_PCA_nmodes}. Red and blue symbols represent the recovered HI $\times$ HI and HI $\times$ ELG, respectively. Different panels correspond to the number of modes removed, as indicated at the top of each panel. Red dashed and blue solid lines correspond to the HI $\times$ HI and HI $\times$ ELG with no foreground, respectively. The recovered power spectra are closest to the ones with no foreground when removing just the first two or three modes. Removing the only first mode from the map leaves too much residual foreground on the sky, and the recovered HI $\times$ HI is over-estimated, with a much larger amplitude than the true spectra. The amplitude of the recovered HI $\times$ ELG cross-correlation is also too large when only one PCA mode is removed, and the spectrum has fluctuations. Conversely, removing the first four modes removes too much of the 21-cm signal, resulting in a significantly under-estimated  power spectrum, compared to the true values. However, the recovered power spectrum are always biased to the true power spectra outside $1\sigma$ due to the over-cleaning in all cases. This could be improved by using more frequency channels in the foreground cleaning procedure, but there would be still the bias issue (see figs 14, 15 \& 16 in \citet{2020arXiv200100833A}). We return to this issue in the next paragraph. Here in terms of the 10 frequency channels, we conclude that the PCA can recover the 21-cm distribution by removing the first two or three modes, otherwise there is a systematic error in the inferred map. The same conclusions are also found using FastICA.

Our analyses below are then based on the reconstructed 21-cm intensity maps, with foregrounds removed using the first two modes only. In the upper panels of Figure~\ref{fig:Cl_PCA_FastICA}, we plot the recovered HI $\times$ HI (red square) and HI $\times$ ELG (blue circle) power spectra, compared with the true ones with no foreground noise (lines). We first present the results without including Tianlai instrument noise in the first two columns, which are based on the PCA and the FastICA, respectively. Both recovered HI $\times$ HI and HI $\times$ ELG are quite close to their corresponding true cases, indicating that most of the foreground noise is already removed from the map. However, there is a systematic bias in both recovered cases. To see the bias level, the normalized difference of $C_\ell$ between the recovered and the true power spectra are shown in the lower panels of Figure~\ref{fig:Cl_PCA_FastICA}. The black dashed lines are shown for the $1\sigma$ error ranges (computed by Eq.~(\ref{eq:Cl_er})) of the true HI $\times$ ELG, similar with the one for HI $\times$ HI. The bias level is significantly out of $1\sigma$ range for both recovered power spectra. This is expected since the foreground and the 21-cm signal is tangled together and there would always be a signal loss in the foreground cleaning procedure. In this case, we would not expect the recovered power spectra to be immune to the bias by the residual foreground. There is also an inescapable bias caused by the signal loss in the cross-correlation case.

What is clear from this analysis, and most importantly from the point of view of cosmological constraints, is that the  pattern of bias introduced by the contamination is different between the recovered autocorrelation and cross-correlation power spectra. We can see clearly that the  offset between the true and measured values is changing with scale in the autocorrelation power spectra, but not in the case of cross-correlation. To quantify the amount of offset as a function of the scale, we then defined a bias factor relating the recovered to the true power spectra by,
\begin{equation}
\label{eq:bias}
     C_\ell^{\rm Recover} = b^2 C_\ell^{\rm True},
\end{equation}
where the true $C_\ell$ gives the power spectra without any noise from the foreground or instrument. The first two rows in Table~\ref{tab:bias_fit_rec} list the best-fitting bias and the corresponding reduced $\chi^2$ when no instrument noise is added. As shown in the cross-correlation case, the reduced $\chi_\nu^2$ values for the best-fitting $b$ are close to unity. This indicates that the recovered cross-correlation power spectra could match the true one using a linear bias. However, in the autocorrelation power spectra case, the reduced $\chi_\nu^2$ values are much larger, indicating that the recovered autocorrelation power spectra cannot be related to the true one by a linear bias, and the offset should depend on the scale.

It seems from these simulations that the recovered HI $\times$ ELG is just linearly biased to the true cross-correlation power spectra. This is a very useful result for the modelling of the cross-correlation power spectra, as we already use a linear galaxy clustering bias as part of the model for that type of data. Thus, the cross-correlation offset caused by the signal loss can be easily parameterized as a linear bias, which can be merged into the linear clustering bias parameter. Note that this would cause the foreground bias and the clustering bias to be degenerate, resulting in the difficulty of measuring the HI bias, but it is useful for modelling the BAO signal, as we discuss in detail in the next section. However, for modelling the recovered HI $\times$ HI autocorrelation power spectra, the scale-dependent bias would require a more complex nuisance parameter.

We expect this result to be generically true, independent of the instrument making the observation. In this paper, to make a realistic prediction, we add the Tianlai instrument thermal noise on the map by assuming two-year observations (see Section \ref{sec:mocking} for details). We repeat reconstruction of the 21-cm distribution using PCA and FastICA, this time including the instrument noise in the map, and measuring the recovered HI $\times$ HI and HI $\times$ ELG power spectrum. Results are shown in the last two columns of Figure~\ref{fig:Cl_PCA_FastICA}. As shown, the HI $\times$ HI auto power spectra show several peaks and troughs and deviate from the true one, which is essentially dominated by the Tianlai instrumental noise that oscillates in the same manner. The baseline distribution of the Tianlai array in the u-v plane is not uniform, leading to the oscillating noise power spectrum. In this case, it would be a difficult challenge to detect any useful clustering information from the autocorrelation power spectra of the 21-cm \citep{2016arXiv160603830Z,2020arXiv200100833A}. In the cross-correlation case, although the recovered HI $\times$ ELG suffers more fluctuations, we could still see its overall clustering pattern, following the true one clearly. This is expected as the instrumental thermal noise is uncorrelated with the galaxies. We also repeat the fitting using Eq.~(\ref{eq:bias}), where the best-fitting bias with its reduced $\chi^2$ is shown in the last two rows of Table~\ref{tab:bias_fit_rec}. As expected, the value of $\chi_\nu^2$ is quite close to unity, indicating that the cross-correlation power spectra still keep scale-independent bias when including the instrument noise in the map. This leads to a positive indication that we could still have modelling of the recovered cross-correlation power spectra by introducing the additional linear bias parameter even though the instrument noise exists.
\begin{table}
    \center
    \caption{SNR, bias fitting and $\chi^2$ per dof for foreground-cleaned map within $700-710$MHz.}
    \label{tab:bias_fit_rec}
    \setlength{\tabcolsep}{3pt}
    \begin{tabular}{lcccccc}
        \toprule
        \multirow{2}{*}{} &
        \multicolumn{3}{c}{\multirow{2}{*}{Auto}} & 
        \multicolumn{3}{c}{\multirow{2}{*}{Cross}} \\\\ 
        
        \multirow{2}{*}{ }  &
        \multirow{2}{*}{$SNR$}  & 
        \multirow{2}{*}{$b$}  & 
        \multirow{2}{*}{$\chi_\nu^2$} &
        \multirow{2}{*}{$SNR$}  & 
        \multirow{2}{*}{$b$}  & 
        \multirow{2}{*}{$\chi_\nu^2$} \\\\
        \hline
        \multirow{2}{*}{PCA (no instru. noise)} & 
        \multirow{2}{*}{$285.03$}  & 
        \multirow{2}{*}{$0.94$}  & 
        \multirow{2}{*}{$93.14$} &
        \multirow{2}{*}{$213.16$}  & 
        \multirow{2}{*}{$0.77$}  & 
        \multirow{2}{*}{$0.73$} \\\\
    
        \multirow{2}{*}{FasctICA (no instru. noise)} & 
        \multirow{2}{*}{$285.04$}  & 
        \multirow{2}{*}{$0.94$}  & 
        \multirow{2}{*}{$93.15$}  &
        \multirow{2}{*}{$213.17$}  & 
        \multirow{2}{*}{$0.77$}  & 
        \multirow{2}{*}{$0.74$}  \\\\ 

        \multirow{2}{*}{PCA (with instru. noise)} & 
        \multirow{2}{*}{$96.70$}  & 
        \multirow{2}{*}{$1.29$}  & 
        \multirow{2}{*}{$1507.2$} &
        \multirow{2}{*}{$85.43$}  & 
        \multirow{2}{*}{$0.76$}  & 
        \multirow{2}{*}{$1.31$} \\\\
    
        \multirow{2}{*}{FasctICA (with instru. noise)} & 
        \multirow{2}{*}{$96.71$}  & 
        \multirow{2}{*}{$1.29$}  & 
        \multirow{2}{*}{$1507.3$}  &
        \multirow{2}{*}{$85.43$}  & 
        \multirow{2}{*}{$0.76$}  & 
        \multirow{2}{*}{$1.32$}  \\\\ 
        \bottomrule
    \end{tabular}
\end{table}
\begin{figure*}
    \centering
    \includegraphics[width=0.9\columnwidth]{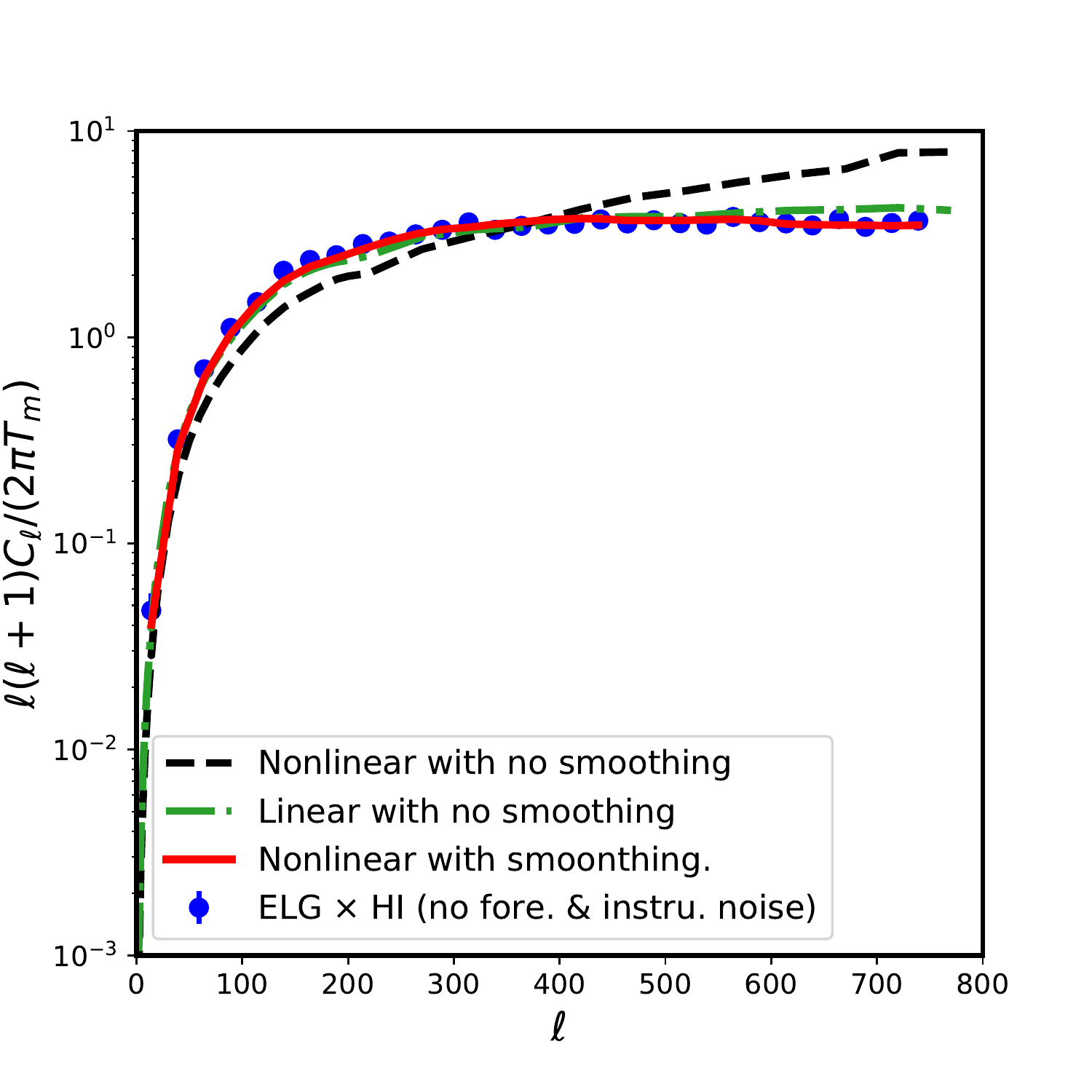}
    \includegraphics[width=0.9\columnwidth]{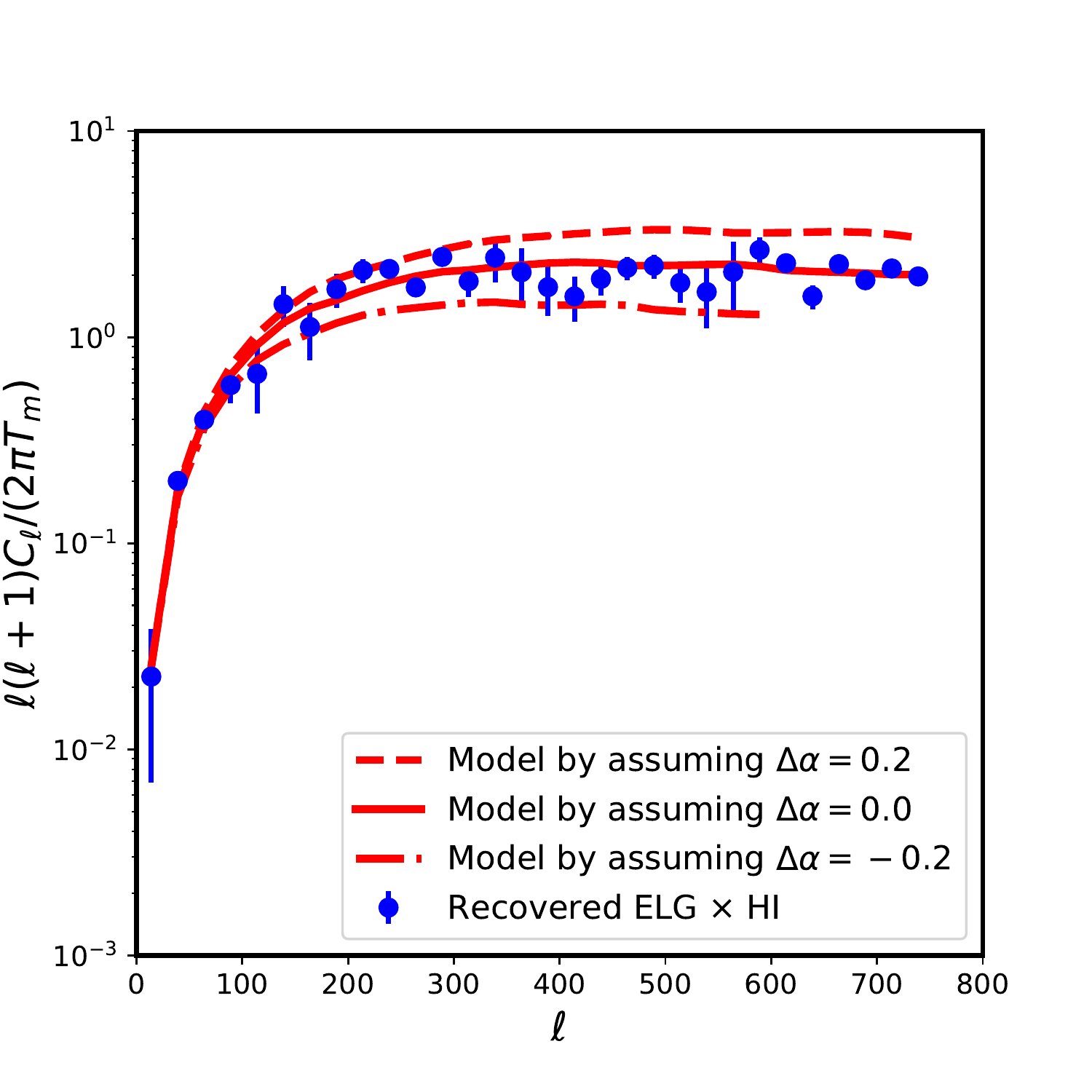}
    \caption{Modelling the angular cross-correlation power spectra HI $\times$ ELG for the map within $700-710$MHz, corresponding to the median redshift $z=1.014$. (Left panel) The blue solid circles are the HI $\times$ ELG measured by using the 21-cm map without adding the foreground and instrument noise. The red solid line is the prediction of the model, using Equation~(\ref{eq:Cl_cros_th}), where the parameters are obtained by fitting to the measured HI $\times$ ELG on scale $50<\ell<767$. For comparison, the prediction of the model without smoothing is also plotted here, where green dash-dot and black dashed lines are for those based on the linear and non-linear power spectrum of dark matter, respectively. (Right panel) We apply the model, Equation~(\ref{eq:Cl_cros_th}) with smoothing procedure, to fit the recovered HI $\times$ ELG with instrument noise added. This is the comparison when assuming different rescale values, $\Delta\alpha$, for $\ell$ according to Equation~(\ref{eq:ell_alpha}). The red solid line is the prediction by assuming $\Delta\alpha=0.0$. For comparison, the red dashed and dash-dotted lines correspond to $\Delta\alpha=-0.2$ and $0.2$, respectively.}
    \label{fig:Cl_model}
\end{figure*}
\begin{figure*}
    \centering
    \includegraphics[width=2.2\columnwidth]{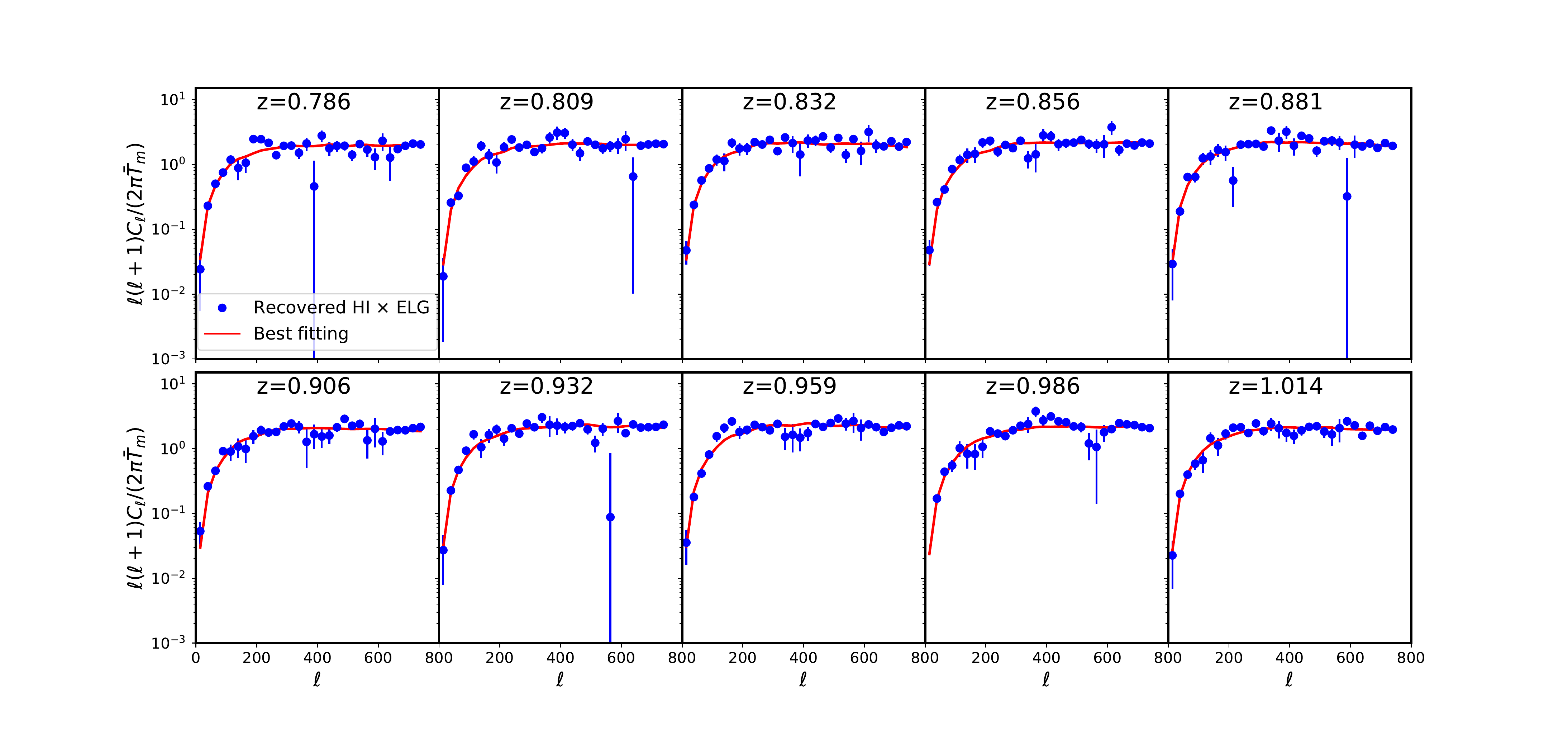}
    \caption{Fitting to the recovered ELG $\times$ HI. Different panels are for the results in 10 redshift bins, as the median redshift of each bin indicated on each panel. The blue circles are the measurements of the angular cross-correlation power spectrum between ELG and the recovered 21-cm, where filled (empty) circles indicate positive (negative) values of Cl. The red lines are the best-fitting model.}
    \label{fig:MCMC_fitCl}
\end{figure*}

\section{Determination of cosmic distance}\label{sec:baofit}

The BAOs are an acoustic peak structure caused by the tension between gravitational infall and outward radiative pressure of the baryon-photon fluid that was imprinted on the large-scale structure at the surface of the last scattering. The peak structure appearing in the correlation function and 3D power spectrum can be (and has already been) exploited to measure cosmic distances, and dubbed a standard ruler of the Universe. These peaks are less visible in the 2D projected power spectrum (when the redshift bins are wide), as modes along the line of sight are integrated to smooth out peak structure. 
Although it is a less model-independent feature to measure cosmic distance, it still provides a useful tool (as shown, for example, in \citet{2012ApJ...761...13S}). This broad-band BAO fitting requires modelling the non-linear clustering and galaxy biases, and marginalising over the common coherent effects in terms of nuisance parameters.  In this section, we explain the appropriate treatment to measure cosmic distance through HI $\times$ ELG cross-correlation.

\subsection{Theoretical formulation for a 2D broad-band power spectrum}
We fit the BAO feature by modelling the angular power spectrum. First, we consider the model for the tracer galaxy as a general case, which is easily applied to other tracers. The angular power spectrum of galaxies can be computed by,
\begin{eqnarray}\label{eq:Cl_gal_th}
    C_\ell^{XY} & = & \frac{2}{\pi}\int{dk k^2 \left[W^X_\ell(k)\right]\left[W^Y_\ell(k)\right]},
\end{eqnarray}
where $X$ and $Y$ denote tracers, such as the 21-cm emitter or ELG galaxy, and the kernel function is,
\begin{eqnarray}\label{eq:b_kernel}
    W^X_\ell(k)
   & = &  \int{dz \phi(z)\sqrt{P_{XX}(k,z)}j_\ell(kr)} \nonumber \\
   & ~ &  \int{dz \phi(z)D(z)\sqrt{P_{XX}(k)}j_\ell(kr)}.
\end{eqnarray}
Here $P_{XX}(k)$ is the 3D power spectrum of tracer $X$,  $r$ is the comoving distance along our past light cone, $\phi(z)$ is the radial selection function, $D(z)$ is the linear growth factor relative to $z=0$ and $b(z)$, is the galaxies bias factor.

Both the 21-cm signal and ELG galaxies trace the underlying density clustering, and the tracing pattern can be formulated using a local bias model, which models the non-linear and non-local halo density bias $\delta_X=\delta_X(\delta_m)$. This functional form is expanded as~\citep{2009JCAP...08..020M},
\begin{eqnarray}
\label{eq:bias_x}
\delta_X(\delta_m)
=b_1\delta+\frac{1}{2}b_2[\delta^2-\sigma_2]+{\rm higher\,\,order\,\, terms}\,,
\end{eqnarray}
where $b_1$ is the linear bias parameter, $b_2$ is the second-order local bias parameter, and the term $\sigma_2$ is introduced to  ensure the condition $\langle \delta_X \rangle = 0$. In this analysis, the higher-order bias model is simplified to ignore non-local bias parts. Then, the power spectrum of both tracers can be written as,
\begin{eqnarray}
P_{XX}(k) 
   & = & \frac{P_{X\delta}^2(k)}{P_{\delta\delta}(k)}  \\
   & = & \frac{\left[b_1^XP_{\delta\delta}(k)+b_2^XP_{b_2,\delta}(k)\right]^2}{P_{\delta\delta}(k)} \nonumber.
\end{eqnarray}
Here $P_{\delta\delta}$ is the autocorrelation power spectra of dark matter density, $\delta$, and $P_{X\delta}$ is the tracers-$\delta$ cross-correlation power spectra. $b_1$ is the linear bias parameter and $b_2$ is the second-order local bias parameter.
Then, Equation~(\ref{eq:b_kernel}) changes to,
\begin{equation}\label{eq:b_kernel2}
W^X_\ell(k) = 
   \int{dz \phi(z)D(z)
   \left[b^X_1(z)\sqrt{P_{\delta\delta}(k)}+b^X_2(z)\sqrt{P_{b_2b_2}(k)}\right]j_\ell(kr)},
\end{equation}
where $P_{b_2b_2}(k) = P^2_{b2,\delta}(k)/P_{\delta\delta}(k)$.
Combined with Equation~(\ref{eq:Cl_gal_th}), we have,
\begin{eqnarray}\label{eq:Cl_cros_th}
    C^{XY}_\ell & = & C^{XY\,b_1}_\ell + C^{XY\,b_2}_\ell
\end{eqnarray}
where,
\begin{eqnarray}
    C^{XY\,b_1}_\ell & = & \frac{2}{\pi}\int{dk k^2 P_{\delta\delta}(k)\left[W^{X\,b_1}_\ell(k)\right]\left[W^{Y\,b_1}_\ell(k)\right]} \nonumber \\
   W^{X\,b_1}_\ell(k) & = &  \int{dz \phi(z)D(z)b^X_1(z)j_\ell(kr)}
\end{eqnarray}
and,
\begin{eqnarray}
    C^{XY\,b_2}_\ell & = & \frac{2}{\pi}\int{dk k^2 P_{b_2b_2}(k)\left[W^{X\,b_2}_\ell(k)\right]\left[W^{Y\,b_2}_\ell(k)\right]} \nonumber \\
   W^{X\,b_2}_\ell(k) & = &  \int{dz \phi(z)D(z)b^X_2(z)j_\ell(kr)}
\end{eqnarray}
Thus, we have a non-linear bias modelling for the angular autocorrelation power spectrum of galaxies. However, when applying this model to the angular cross-correlation power spectrum between HI and galaxies, we have an additional procedure that needs to be considered. As the 21-cm observation is used to measure the aggregate emission from many unresolved galaxies, we need to include a smoothing window function in Equation~(\ref{eq:Cl_cros_th}). We then multiply $C^{XY\,b_1}_\ell$ and $C^{XY\,b_2}_\ell$ by a Gaussian smoothing kernel $W_s(\ell\theta_{\rm sm}) = {\rm exp}\left[-(\ell\theta_{\rm sm})^2/2\right]$, where $\theta_{\rm sm}$ is the smoothing angle radius. As we are working with pixelized HI and galaxy data, in the form of HEALPIX maps \citep{2005ApJ...622..759G}, it is highly likely that this smoothing scale will be related to the pixelization scale.

In the left panel of Figure~\ref{fig:Cl_model}, we make a comparison of HI $\times$ ELG between the simulated measurement and this model. Note that, as discussed below, the angular power spectra are normalized to the mean temperature of the reconstructed 21-cm maps. The blue circles are the measurements obtained by using the 21-cm map without adding the foreground and instrument noise. The red solid line is the prediction of the model, by Equation~(\ref{eq:Cl_cros_th}), which is fitted to the measurement on scale $50<\ell<767$. As shown, they have very good agreement with each other from large to small scale. To show the performance of the smoothing, we also plot the predictions of the model based on the linear (green dash-dotted) and non-linear (black dashed) power spectrum of dark matter without smoothing. The non-linear power spectrum fails to make a prediction on both large and small scales. Although the linear power spectrum recovers well on large scales, it still overestimates the values on a small scale. Therefore, in our paper, we apply the model based on the non-linear dark matter power spectrum with the additional smoothing procedure to fit the cross-correlation power spectra HI $\times$ ELG, and focus on the cosmic distance constraint.   
\begin{figure*}
    \centering
    \includegraphics[width=1.3\columnwidth]{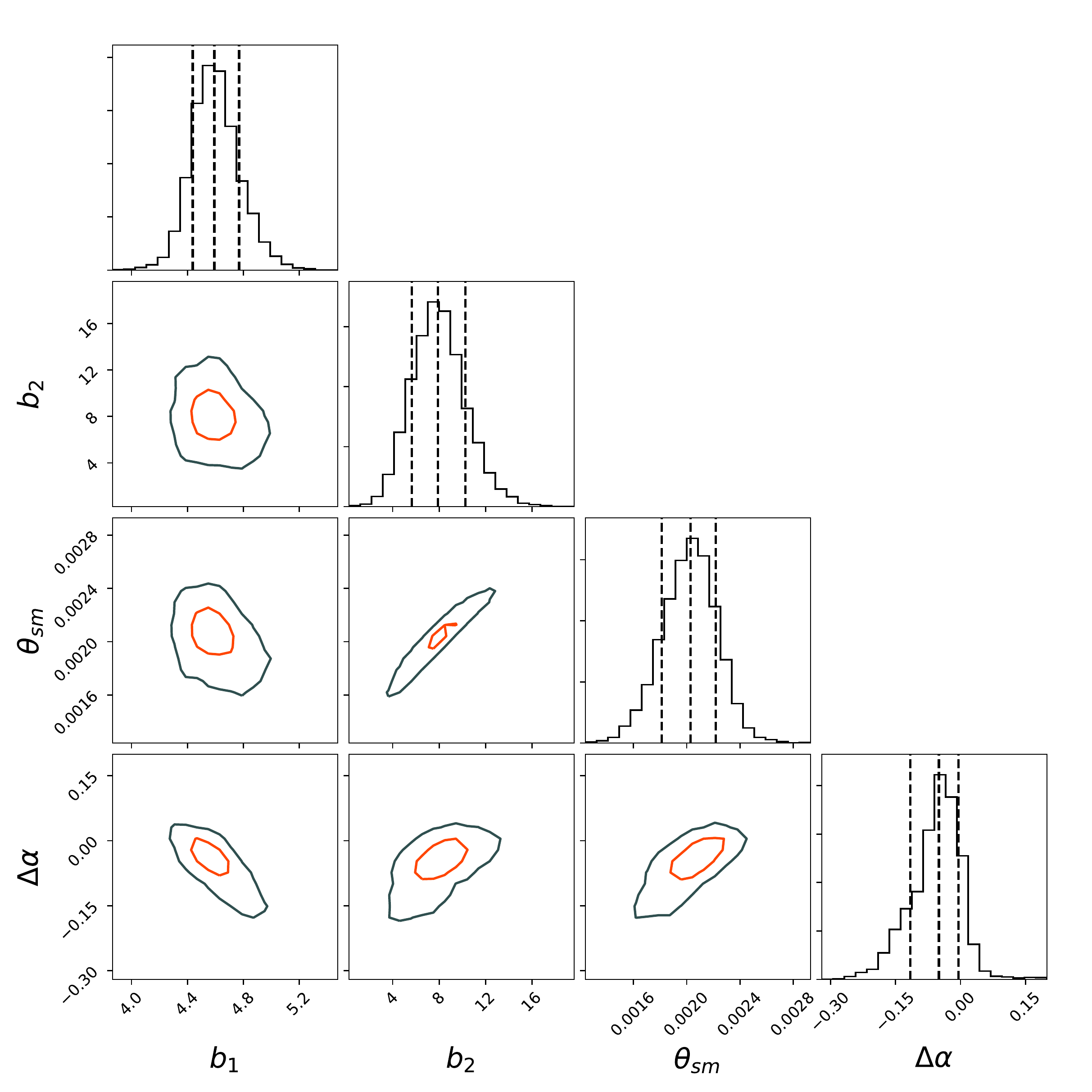}
    \caption{Best-fitting and projected distribution of the parameters in two- or one-dimensional space for the redshift bin $z=0.786$. The red and black contours in the two-dimensional plane correspond to the boundaries of $68\%$ and $95\%$ confidence levels, respectively. The one-dimensional distributions are the marginalized distributions of individual parameters. The vertical black lines indicate the best-fitting values and the 68\% confidence region.}
    \label{fig:MCMC_contour}
\end{figure*}
\begin{figure}
    \centering
    \includegraphics[width=1.1\columnwidth]{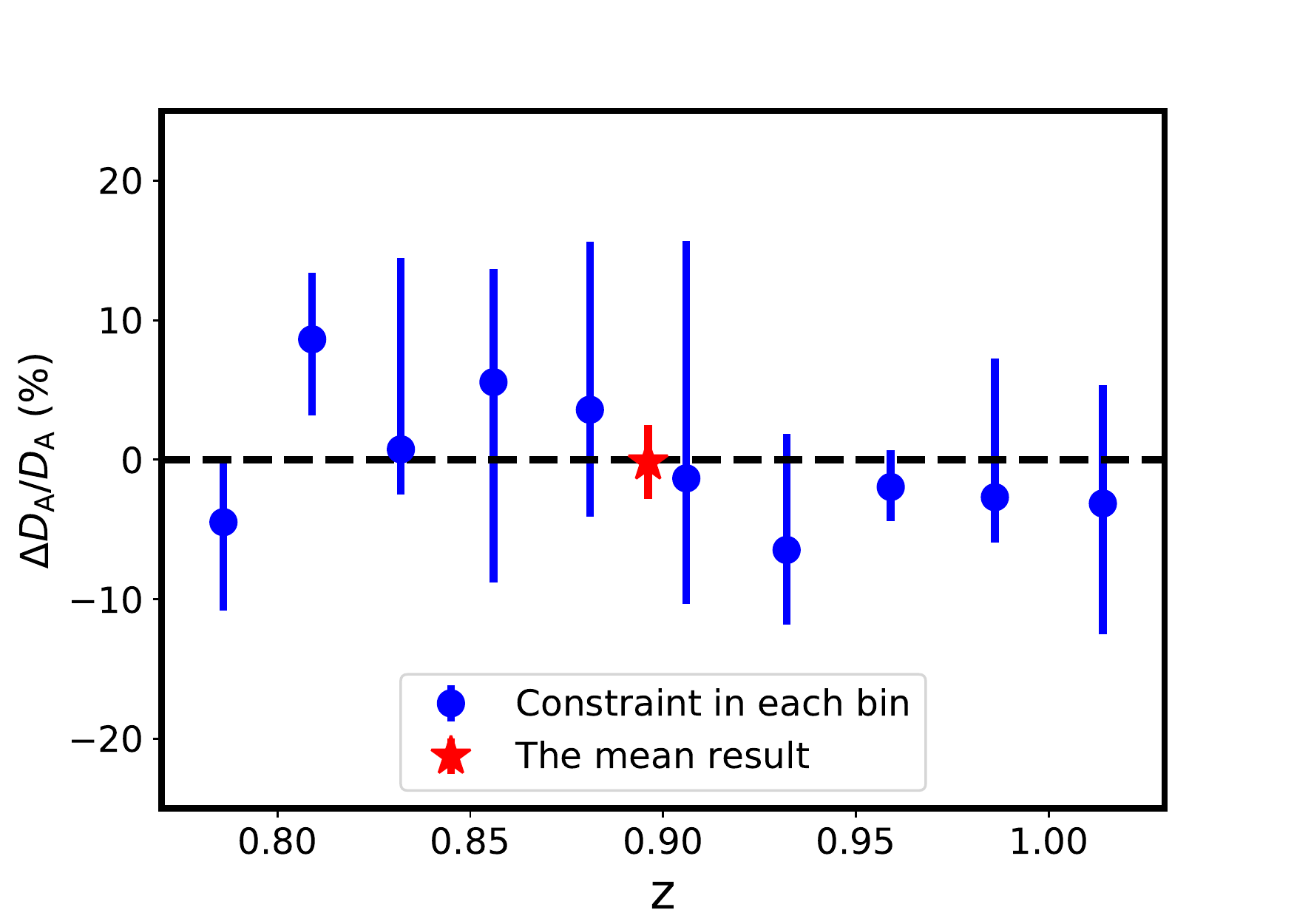}
    \caption{Best-fitting $\Delta \alpha$ (as a percentage of the true value) using the MCMC method. The blue circles show the results for each of the 10 redshift bins, with error bars indicating the 68\% confidence regions. The red star show the mean result averaged from the 10 redshift bins.}
    \label{fig:MCMC_DAerr}
\end{figure}
\begin{figure}
    \centering
    \includegraphics[width=1.1\columnwidth]{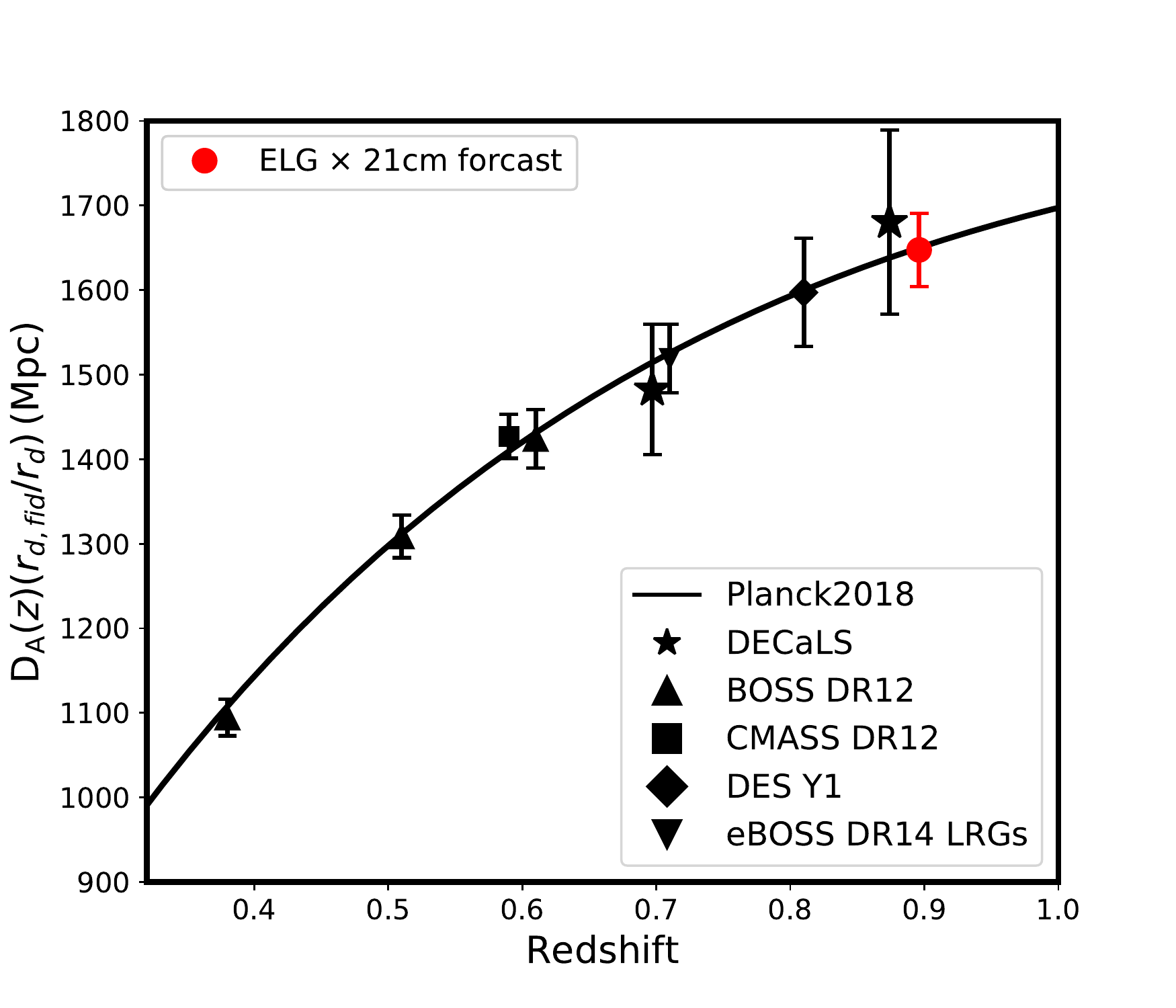}
    \caption{Comparison of the $D_{A}(z)$ measurements between the present results from galaxy surveys and the forecast based on the ELG $\times$ 21-cm cross-correlation. The red circle is the forecast from the angular cross power spectrum between DESI ELG and Tianlai 21-cm in the Tianlai Pathfinder redshift range $0.78<z<1.03$. The black symbols are measurements by other surveys, as indicated in the legend. The solid black line corresponds to the theoretical predictions for $D_{A}(z)$ as a function of redshift obtained using the cosmological parameters measured from \citep{Planck_2018}. }
    \label{fig:DA_z}
\end{figure}
\subsection{Probing broad-band BAO}

The BAO in the radial and tangential directions provide measurements of the Hubble parameter and angular diameter distance, respectively. Here we focus on using the angular power spectrum to constrain the angular diameter distance. The subtended angle, $\theta(z)$, of the sound horizon length is determined by, 
\begin{equation}
    \theta = \frac{r_s(z_{\rm drag})}{(1+z)D_\rmA(z)}\,,
\end{equation}
where $D_\rmA(z)$ is the proper angular diameter distance, and $r_s(z_{\rm drag})$ is the sound horizon at the drag epoch, which we fix as the fiducial value. The length-scale of the multipoles of degree, $\ell$, in the angular power spectrum is related to the angular diameter distance by, 
\begin{equation}\label{eq:ell_DA}
    \ell \propto \frac{1}{\theta} = \frac{(1+z)D_\rmA(z)}{r_{\rm s}(z_{\rm drag})}.
\end{equation}
Consequently, assuming an incorrect $D_\rmA(z)$ can result in systematic errors in the multipoles of degree, $\ell$, which turn out to shrink the shape of the angular power spectrum. We can then model such shrinking deviations and constrain $D_\rmA(z)$ by searching for the model that matches the measurement best. Then, we define the rescaling parameter as,
\begin{equation}\label{eq:ell_alpha}
    (1+\Delta\alpha)\ell \propto (1+\Delta\alpha)D_\rmA(z)\,,
\end{equation}
where we rescale $D_\rmA(z)$ by a factor $1+\Delta\alpha$ to introduce the error coming from the changes of $D_\rmA(z)$ into the $C_\ell$. As $\ell$ is directly proportional to $D_\rmA(z)$, $\ell$ is also changed by the same rescaling amount. We then derive constraints on $D_\rmA(z)$ by searching for the best-fitting $\Delta\alpha$.
 In order to see how the angular power spectrum is affected by the rescaling parameter $\Delta\alpha$, we apply $\Delta\alpha=0.2$, $0.0$ \& $-0.2$ in the models, which are plotted with red, black and green lines respectively in Figure~\ref{fig:Cl_model}. It is reassuring that the theoretical $C_\ell$ best matches the measurement for $\Delta\alpha=0$, which is the value that corresponds to the fiducial $D_\rmA(z)$ in the simulation. Adopting $\Delta\alpha=0.2$ or $-0.2$, corresponding to a 20 per cent difference with the true $D_\rmA(z)$, results in an inferred bias for the $C_\ell$ that is systematically too high or too low.

We now derive constraints on the related parameters, $\Delta\alpha$, as well as the bias parameters, $b_1$, $b_2$ and $\theta_{\rm sm}$. The biases are defined as,
\begin{eqnarray}
\label{eq:b_coherent}
b_1&\equiv& b_{\rm fg}g_\delta\sqrt{b^{\rm{HI}}_1b_1^{\rm{ELG}}}\nonumber \\
b_2&\equiv& g_\delta\sqrt{b^{\rm{HI}}_2b^{\rm{ELG}}_2}\ \,,
\end{eqnarray}
where $b_{\rm fg}$ denotes the coherent offset caused by information loss in foreground cleaning procedure, and $g_\delta$ denotes the growth factor of particle clustering. Here we do not separately fit $(b_{\rm fg}, b^{X}_1, b^{X}_2,g_\delta)$, because all are coherent parameters and are not distinguishable. To keep the clustering at the same level in different redshifts, we normalized the angular power spectrum to the mean temperature of the reconstructed 21-cm maps in the whole Tianlai redshift range.  We then use a Markov chain Monte Carlo (MCMC) method to explore the likelihood function in the multidimensional space. We use the affine-invariant ensemble sampler, known as the MCMC  hammer and  described  in \citet{2013PASP..125..306F}\footnote{The  code  emcee  can  be  found  at \url{https://github.com/dfm/emcee}.} to sample the parameter space.
The corresponding $\chi^2$ is defined as,
%
\begin{equation}
    \chi^2 = {\sum_{\ell}{\frac
    {\left(C^{\rm mod}_{\ell(1+\Delta\alpha)}(z)-
     C^{\rm mea}_{\ell}(z)/\bar{T}_{m}\right)^2}
    {\sigma^2_{\ell}(z)}}},
\end{equation}
where $C^{\rm mea}_{\ell}(z)$ is the power spectrum measured from the map at redshift $z$, with $\sigma_{\ell}(z)$ for the error obtained from Equation~(\ref{eq:Cl_er}), $C^{\rm mod}_{\ell(1+\Delta\alpha)}(z)$ is the model power spectrum computed by using Equation~(\ref{eq:Cl_cros_th}), with $\ell$ multiplied by the factor $1+\Delta(\alpha)$, and $\bar{T}_{m}$ is mean temperature of the reconstructed 21-cm intensity map in the Tianlai redshift range.

To show the agreement of fit between the model and simulated sky, Figure~\ref{fig:MCMC_fitCl} shows the comparison between the MCMC best-fitting models (red line) and the measurements (blue circles) for the 10 redshift bins, at the median redshift of each bin indicated on each panel. We can see that they agree quite well with each other. Figure \ref{fig:MCMC_contour} shows the projected  two-dimensional confidence regions in the parameter space for the first redshift bin, to which other bins are quite similar. The red and black contours indicate the $68\%$ and $95\%$ confidence levels, respectively. In the top panels, it shows also the marginalized, one-dimensional distribution for each parameter, with vertical dashed lines indicating the mean and $68\%$ confidence regions. 

In terms of precision of the distance measurements, figure \ref{fig:MCMC_DAerr} shows the best-fitting $\Delta\alpha (\%)$ for each of the 10 redshift bins, with error bars indicating the 68\% confidence intervals. As can be seen, the results are in agreement with the expected values within the $1\sigma$ level. However, the results  not only suffer from the precision error introduced by the residual foreground and instrument noise, but also the accuracy problem from cosmic variance in different redshift slices. Finally, we forecast a result for the whole redshift bin by averaging the results from the 10 bins, as shown with the red star in Figure \ref{fig:MCMC_DAerr}. Here we calculate the mean result by only considering the diagonal term of the covariance matrix, under the assumption that the measurements in different redshift bins are independent. In this case, our mean results imply that $\Delta\alpha=-0.002 \pm 0.027$, with the error calculated by the standard deviation of the mean formula,
\begin{equation}
    \sigma = \frac{\sqrt{\sum^{10}_i\sigma^2_{\rm i}}}{10},
\end{equation}
where $\sigma_{\rm i}$ is the MCMC error in each of the 10 redshift bins. The mean result indicates that the constraint on $D_\rmA$ can be at the 2.7\% level. To obtain a realistic estimation, we need to calculate the error by using the full covariance matrix information, which can be obtained by generating many different realizations.

Moreover, in order to show the ability of constraint from the cross-correlation, in Figure \ref{fig:DA_z}, we compare our forecast on $D_\rmA$ to the measurements from previous galaxy surveys, based on the galaxy clustering only. These include the results from BOSS DR12 \citep{Alam_2017}, CMASS DR12 \citep{Chuang_2017}, eBOSS DR14 \citep{eBOSS_2018}, DES photometric redshift survey \citep{DES_Y1}, and DECam Legacy Survey (DECaLS) \citep{2020arXiv200513126S}. The solid black line corresponds to the theoretical predictions as a function of redshift obtained using the cosmological parameters from \citep{Planck_2018}. We then conclude that, by using the angular cross-correlation power spectrum between the ELG and the recovered HI, we would be able to put successful constraints on the angular diameter distance $D_\rmA$ to a good level in the whole redshift bin. 

\section{Conclusions}
\label{sec:conclu}

Using the HI autocorrelation power spectra to derive cosmological constraints will be challenging, as  non-trivial residuals from foreground cleaning and the effect of the instrument noise will both be difficult to model, as presented in detail by \citet{2020arXiv200100833A}. In this follow-on manuscript, we find that the foreground cleaning residual exhibits a coherent effect on the measured cross-correlation spectra, and the instrument noise is removed by the cross-correlation between different tracers observed at separate instruments. These two important effects allow us to accurately model the broad-band shape of the underlying density clustering at the precision level.

We show how these features can be applied to probe the BAOs at the HI source redshift, and we verified that the cosmic distances will be measurable at high precision. This idea has been roughly outlined in many previous works in the literature, but this is the first time it has been rigorously demonstrated through simulations. We have also presented accurate modelling at the percentage precision level, along with a detailed theoretical strategy to probe BAO that includes the combined bias treatment. We show that a precise and accurate cosmological constraint is achievable with relatively poor resolution HI experiments, such as Tianlai. While is certainly true that similar constraints will be made exploiting galaxy autocorrelations alone, the alternative measurements with various tracers are necessary to confirm the results,  and exclude the possibility that the measurements are contaminated by systematic uncertainties of each individual tracer.

Moreover, we made detailed analyses on the foreground-cleaned map. As information is lost in the cleaning procedure, there is an inescapable bias in the recovered power spectrum. However, we have found that the bias from the cross-correlation power spectrum between HI and ELG is scale-independent, while the offset in the HI auto-correlation power spectrum is scale-dependent. The scale-independent bias is a useful feature in that it can be easily parameterized as a linear bias in the modelling of the cross-correlation power spectrum. We have developed a method to model the angular cross-correlation power spectrum by considering the coherent bias in the linear scheme as defined by Eq.~(\ref{eq:b_coherent}).

We applied this model of cross-correlation bias to fit the broad-band BAO feature measured from the angular cross-correlation power spectra between ELG and the recovered HI. We forecast a constraint on the angular diameter distance with a precision of 2.7\% at the Tianlai redshift range $0.775<z<1.03$. Considering the challenge of the Tianlai instrument noise, it is encouraging to see such a measured cosmic distance at several percentage precision levels using alternative tracer, which will be complementary to the BAO cosmic distance measured by galaxy-galaxy auto-correlation. In the next-generation galaxy surveys, an accuracy of 1 per cent is required in determining the ``standard ruler'' of large-scale structure of the Universe. Future HI experiments, such as the SKA, will provide the 21-cm intensity mapping with improved instrument noise, along with larger volume and higher resolution. It is expected that the cross-check on cosmic distance will be enhanced by the cross-correlation between the galaxy and the HI.

It is still a challenge to constrain the full cosmology from the 21-cm intensity map. The cross-correlation is immune to the error by the residual foreground and the instrument noise in terms of the amplitude, but it still suffers from the linear offset due to the signal loss by the over-cleaning. This would be not an issue in detecting the BAO signal where we can make a fitting with the coherent bias, but it will need to be corrected in constraining the HI bias, the HI abundance, and the growth rate. However, it is hard to make such correction because the real HI distribution is unknown. One possible way is to define a transfer function to quantify the over cleaning by running the foreground cleaning on the pure 21-cm intensity map, and so to calibrate the amplitude correction \citep[e.g.][]{2020arXiv200100833A}.

\section*{Acknowledgements}

We would like to thank Sungwook Hong, Benjamin L’Huillier, Juhan Kim, and Changbom  Park  for  multiple  discussions  and  for  providing  the  HR4  simulations. We would also like to thank Xuelei Chen and Yidong Xu for providing us details on the Tianlai instrument. F. Shi would like to thank Jiajun Zhang and Fei Qin for useful discussions. F. Shi,  Y-S.  Song, J.  Asorey, and D.  Parkinson   are  supported  by  the  project \begin{CJK}{UTF8}{mj}우주거대구조를 이용한 암흑우주 연구\end{CJK} (“Understanding  Dark  Universe Using Large Scale Structure of the Universe”), funded by the Ministry of Science. K. Ahn is supported by NRF-2016R1D1A1B04935414  and  NRF-2016R1A5A1013277. L. Zhang is supported by the National Key R\&D Program of China (2018YFA0404504, 2018YFA0404601), the National Science Foundation of China (11621303, 11890691, 11653003, 11773021), the 111 project, and the CAS Interdisciplinary Innovation Team (JCTD- 2019-05). 

This research made use of Astropy,\footnote{\url{http://www.astropy.org}} a community-developed core Python package for Astronomy \citep{astropy:2013, astropy:2018}, the HEALPix and Healpy package \citep{2005ApJ...622..759G,Zonca2019}, the  Numpy package \cite{book}, the Scipy package \citep{2019arXiv190710121V} and Matplotlib package \citep{Hunter:2007}.

\section*{Data availability}

The data underlying this article will be shared on reasonable request to the corresponding author.




\bibliographystyle{mnras}
\bibliography{reference} 

\bsp	
\label{lastpage}
\end{document}